\documentclass[11pt,number,preprint,3p]{elsarticle}
\usepackage{titlesec}
\usepackage{color}
\usepackage[dvipsnames]{xcolor}
\usepackage{soul}
\usepackage{graphicx}
\usepackage{subcaption}
\usepackage[colorlinks]{hyperref}
\makeatletter
\gdef\urlauthor#1#2{\g@addto@macro\@elsuads{\let\corref\@gobble%
     \def\@@tmp{#1}\raggedright\eadsep
     {\ttfamily\url{\expandafter\strip@prefix\meaning\@@tmp}}\space(#2)%
     \def\eadsep{\unskip,\space}}%
}
\gdef\emailauthor#1#2{\stepcounter{ead}%
     \g@addto@macro\@elseads{\raggedright%
      \let\corref\@gobble\def\@@tmp{#1}%
      \eadsep{\ttfamily\href{mailto:\expandafter\strip@prefix\meaning\@@tmp}{\expandafter\strip@prefix\meaning\@@tmp}}
      (#2)\def\eadsep{\unskip,\space}}%
}
\makeatother
\usepackage{amssymb}
\usepackage{amsthm}
\usepackage{amsmath}
\usepackage{amssymb}
\usepackage{mathtools}
\usepackage{dsfont}
\usepackage{booktabs}
\usepackage{url}
\usepackage{scrextend}
\usepackage{epstopdf}
\usepackage{float}
\usepackage{tcolorbox}
\usepackage{tablefootnote}
\usepackage[perpage]{footmisc}
\usepackage{lineno}
\usepackage{multirow}

\usepackage[ruled,vlined]{algorithm2e}

\newcommand{\vect}[1]{\boldsymbol{#1}}



\makeatletter
\let\@afterindenttrue\@afterindentfalse
\makeatother

\biboptions{numbers,sort&compress,square}

\usepackage{enumitem}

\setlist[itemize]{itemsep=3pt, topsep=3pt, parsep=3pt, partopsep=3pt}
\setlist[enumerate]{itemsep=3pt, topsep=3pt, parsep=3pt, partopsep=3pt}

\begin{document}

	\begin{frontmatter}
		\renewcommand{\thefootnote}{\fnsymbol{footnote}}
		\title{First-passage reliability sensitivity analysis of linear systems subjected to non-Gaussian wind excitations by surface decomposition method}
		\author[1,2]{Jianhua Xian}  
            \author[2]{Sai Hung Cheung\corref{cor1}}
            \ead{jhcheung@hku.hk}
            \cortext[cor1]{Corresponding author}
            \author[1,3,4]{Cheng Su}     
		\address[1]{School of Civil Engineering and Transportation, South China University of Technology, Guangzhou, China}
        \address[2]{Department of Civil Engineering, The University of Hong Kong, Hong Kong, China}
        \address[3]{State Key Laboratory of Subtropical Building and Urban Science, South China University of Technology, Guangzhou, China}
        \address[4]{School of Civil Engineering, Guangzhou City University of Technology, Guangzhou, China}

	\begin{abstract}
        This contribution develops a surface decomposition method for first-passage dynamic reliability sensitivity analysis of linear systems exposed to non-Gaussian wind excitations. The first-passage failure probability sensitivity is formulated as a system surface integral over a highly non-smooth and high-dimensional hypersurface. This complex integral is first decomposed into a collection of component surface integrals over the truncated smooth quadratic hypersurfaces. The dominant components are then identified based on the relative magnitudes of the first-order approximations of the component failure probabilities. A nested sampling algorithm is constructed to efficiently estimate the sum of these component surface integrals, in which the number of system limit-state function evaluations equals the number of outer-level samples while remaining independent of the inner-level sample size. A key advantage of the present approach is that the function evaluation results can be reused across different design parameters. Two numerical examples are explored to demonstrate the effectiveness of the proposed method. The results indicate that the number of function evaluations required is typically below 100 to achieve a target coefficient of variation of 0.1. 
        \end{abstract}
  
    
	\end{frontmatter}

	\renewcommand{\thefootnote}{\fnsymbol{footnote}}
	
\section{Introduction}

External loads acting on engineering structures often exhibit significant non-Gaussian statistical characteristics, such as earthquake loads \cite{radu2018site}, wind loads \cite{gurley1997analysis}, wave loads \cite{hu1987non}, and moving vehicular loads \cite{chen2006modeling}. For the convenience of statistical description and subsequent reliability analysis, these external loads are commonly modeled as Gaussian random processes. However, such a Gaussian assumption may fail to adequately capture the non-Gaussian features of actual excitations and can consequently lead to inaccurate predictions of structural responses and reliability \cite{grigoriu1984crossings,lutes1984stochastic}. For instance, neglecting the quadratic components of non-Gaussian wind excitations can result in a significant overestimation of structural safety \cite{katafygiotis2009reliability}. Therefore, it is of great importance to account for the non-Gaussian nature of random excitations in structural reliability analysis. 

Various effective methods have been developed for the first-passage dynamic reliability analysis of structures exposed to Gaussian random excitations \cite{au2001first,katafygiotis2006domain,xian2024relaxation,xian2024physics}. In contrast, the corresponding research under non-Gaussian random excitations is relatively less developed, although increasing attention has been devoted to this topic in recent years. In particular, Wang et al. \cite{Wang2016efficient} developed the domain decomposition method for the first-excursion dynamic reliability problems of linear structures subjected to non-Gaussian wind loads. Psaros et al. \cite{psaros2018wiener} investigated the Wiener path integral technique for determining the evolutionary probability density functions of responses of marine structures excited by non-Gaussian wave loads. Xu et al. \cite{xu2022dynamic} extended the probability density evolution method for the first-passage dynamic reliability analysis of nonlinear structures driven by non-Gaussian seismic loads. Zhang et al. \cite{zhang2025directional} investigated the directional importance sampling for the first-excursion dynamic reliability analysis of linear structures exposed to non-Gaussian white noise excitations. In addition, recent advances in the random vibration analysis of linear and nonlinear structures under non-Gaussian excitations can be found in \cite{xian2023non,xian2026explicit}.

Reliability sensitivity analysis characterizes the dependence of system reliability on the system parameters and provides quantitative measures of their impacts on the system failure probability. The resulting sensitivity information is crucial for risk-informed decision-making \cite{song2009system,taflanidis2011simulation} and reliability-based design optimization \cite{chun2016structural,yang2022structural,li2023direct,xian2023reliability}. Early studies primarily focused on the sensitivity analysis of static reliability, where the sensitivity estimation methods were developed by extending existing reliability analysis techniques, including sampling methods \cite{lu2008reliability,song2009subset,papaioannou2018reliability}, surrogate modeling methods \cite{dubourg2014meta,torii2017probability,yun2020adaptive}, and moment methods \cite{lu2010reliability}. In recent years, increasing research efforts have been devoted to the sensitivity analysis of first-passage dynamic reliability, which presents significantly greater challenges than its static counterpart. For instance, Chun et al. \cite{chun2015parameter} developed a sequential compounding method to compute parametric sensitivities of the system failure probability, and the method was validated using a first-passage dynamic reliability problem of a linear system under Gaussian excitation. Jensen et al. \cite{jensen2015reliability} presented a subset-simulation-based approach for the sensitivity analysis of first-excursion failure probability for nonlinear structures subjected to Gaussian seismic excitations. Valdebenito et al. \cite{valdebenito2021sensitivity} investigated the first-passage failure probability sensitivities of linear structures excited by Gaussian loadings using directional importance sampling. Misraji et al. \cite{misraji2026first} extended the multi-domain line sampling for sensitivity analysis of first-excursion failure probability of linear systems subjected to Gaussian excitations. Au and Cao \cite{Au2026reliability} presented a Monte Carlo strategy with kernel smoothing for the reliability sensitivity estimation, and the method was applied to the first-passage reliability sensitivity analysis of a linear system under Gaussian white noise. Despite the above advances, the research on the sensitivity analysis of first-passage dynamic reliability of structures under non-Gaussian random excitations remains limited.

Recently, an effective surface decomposition method has been proposed for the first-passage reliability sensitivity analysis of linear structures subjected to Gaussian random excitations \cite{xian2026surface}. The key idea is to decompose the failure probability sensitivity into a sum of component surface integrals over the truncated smooth hyperplanes, which are substantially easier to evaluate than the original system surface integral defined over the highly non-smooth hypersurface. Building upon this idea, the present study further extends the surface decomposition method to reliability sensitivity analysis considering non-Gaussian wind excitations. Since the component limit-state hypersurfaces associated with non-Gaussian wind excitations are quadratic rather than planar, the failure probability sensitivity is decomposed into a collection of component surface integrals over the truncated smooth quadratic hypersurfaces. The dominant component surface integrals are identified through importance sampling based on the relative magnitudes of the first-order approximations of component failure probabilities, in contrast to the original method that relies on the exact component failure probabilities. A two-stage nested sampling algorithm is proposed to estimate the total contribution of these component surface integrals, in which the number of system limit-state function evaluations is determined solely by the number of outer-level samples. The effectiveness of the proposed method is demonstrated through performing the first-passage dynamic reliability sensitivity analyses of a linear oscillator and a 20-story planar braced frame structure under non-Gaussian wind excitations.

\section{Explicit formulation of dynamic response and sensitivity considering non-Gaussian wind excitations}
\subsection{Non-Gaussian wind excitations}
For the along-wind excitations, the wind velocity field can be discretized into a Gaussian random vector process, with each wind velocity process expressed as
\begin{equation}\label{wind velocity}
    V_j(t) = \bar{V}_j +v_j(t) \quad (j=1,2,\dots,N_v)
\end{equation}
where $V_j(t)\;(j=1,2,\dots,N_v)$ is the $j$th wind velocity process at the height $h_j$, with $N_v$ being the number of discretized heights; $\bar{V}_j$ is the mean component of $V_j(t)$; $v_j(t)$ is the fluctuating component of $V_j(t)$, which is assumed to be a zero-mean Gaussian random process.

The wind load induced by the $j$th wind velocity process $V_j(t)$ is expressed as \cite{simiu2019wind}
\begin{equation}\label{wind load1}
    F_j(t) = \frac{1}{2}\rho C_d S_jV_j^2 (t)
     \quad (j=1,2,\dots,N_v)
\end{equation}
where $\rho=1.25\mathrm{kg/m^3}$ is the air density; $C_d$ is the drag coefficient (taken as 1 in this study); $S_j$ is the effective area subjected to the wind pressure induced by $V_j(t)$.

Substitution of Eq.\eqref{wind velocity} into Eq.\eqref{wind load1} yields
\begin{equation}\label{wind load2}
    F_j(t) = \bar{F}_j+\lambda _{j,1}v_j(t)+\lambda _{j,2}v_j^2(t) \quad (j=1,2,\dots,N_v)
\end{equation}
where $\bar{F}_j=\rho C_dS_j\bar{V}_j^2/2$ represents the static wind load caused by the mean wind speed $\bar{V}_j$; and $\lambda _{j,1}=\rho C_d S_j\bar{V}_j$ and $\lambda _{j,2}=\rho C_d S_j/2$ are the coefficients associated with the linear and quadratic components of the buffeting force induced by the $j$th fluctuating wind velocity $v_j(t)$, respectively. 

It can be seen from Eq.\eqref{wind load2} that the wind load acting on the structure is a quadratic function of a Gaussian random process, exhibiting typical non-Gaussian characteristics. It is noted that the quadratic term is often neglected in engineering practice for simplicity. However, it has been found in \cite{katafygiotis2009reliability} that omitting this quadratic term may lead to a significant underestimation of the first-passage failure probability and therefore should be retained.

The zero-mean Gaussian wind velocity vector process $\vect v(t) = [v_1(t)\; v_2(t)\; \cdots\; v_{N_v}(t)]^{\mathrm{T}}$ is fully characterized by its cross-power spectrum matrix $\vect S_{\vect v}(\omega,t)$. Based on the spectral representation method, the wind velocity $v_j(t)$ can be expressed as \cite{deodatis2025spectral}
\begin{equation}\label{spectral representation}
    v_j(t) = \sqrt{2\Delta \omega} \sum_{l=1}^{N_v} \sum_{k=1}^{N_\omega}  
H_{jl}(\omega_k,t)\Big(x_{lk,1} \cos(\omega_k t) + x_{lk,2} \sin(\omega_k t) \Big) \quad (j=1,2,\dots,N_v)
\end{equation}
where $N_\omega$ is the number of circular frequency intervals; $\Delta \omega=(\omega_{\max}-\omega_{\min})/N_\omega$ is the width of each interval, with $\omega_{\max}$ and $\omega_{\min}$ denoting the upper and lower cutoff circular frequencies, respectively; $\omega_k=\omega_{\min}+(k-0.5)\Delta \omega \; (k=1,2,\dots,N_\omega)$ is the circular frequency at the center of the $k$th interval; $H_{jl}(\omega_k,t)$ is the element located at the $j$th row and $l$th column of the matrix $\vect H(\omega_k,t)$, which is the lower triangular matrix obtained through the Cholesky decomposition of $\vect S_{\vect v}(\omega_k,t)$; $x_{lk,1}$ and $x_{lk,2}\;(l=1,2,\dots,N_v;k=1,2,\dots,N_\omega)$ are the mutually independent standard Gaussian random variables.

Eq.\eqref{spectral representation} can be rewritten in the following compact form as
\begin{equation}\label{spectral representation1}
    v_j(\vect X,t) = \vect \psi_j(t) \vect X \quad (j=1,2,\dots,N_v)
\end{equation}
where $\vect X$ is a $d$-dimensional column vector that collects all standard Gaussian random variables $x_{lk,1}$ and $x_{lk,2}\;(l=1,2,\dots,N_v;k=1,2,\dots,N_\omega)$; $d=2N_v N_\omega$; and $\vect \psi_j(t)$ is a $d$-dimensional row vector comprising the deterministic functions $H_{jl}(\omega_k,t)$, $\cos(\omega_k t)$ and $\sin(\omega_k t)$. The explicit construction of $\vect \psi_j(t)$ is straightforward but tedious, and thus is omitted for brevity.

\subsection{Explicit expressions of dynamic responses}
For a linear system subjected to non-Gaussian random wind excitations, the equation of motion can be expressed as
\begin{equation}\label{motion equation}
    \vect M \ddot{\vect U}(\vect X ,t)+\vect C \dot{\vect U}(\vect X ,t)+\vect K \vect U(\vect X ,t)
    = \sum_{j=1}^{N_v}\vect L_j  F_j(\vect X ,t) 
\end{equation}
where $\vect M$, $\vect C$ and $\vect K$ are the mass, damping and stiffness matrices of the linear system, respectively; $\ddot{\vect U}(\vect X ,t)$, $\dot{\vect U}(\vect X ,t)$ and $\vect U(\vect X ,t)$ are the acceleration, velocity and displacement vectors of the linear system, respectively; $F_j(\vect X ,t)$ is the $j$th non-Gaussian random wind excitation written as 
\begin{equation}\label{wind load3}
    F_j(\vect X,t) =\bar{F}_j+\lambda _{j,1}v_j(\vect X,t)+\lambda _{j,2}v_j^2(\vect X,t) \quad (j=1,2,\dots,N_v)
\end{equation}
and $\vect L_j$ is the corresponding orientation vector.

Based on the linear superposition principle, any response $r(\vect X ,t)$ of the linear system can be written in the following form:
\begin{equation}\label{superposition}
    r(\vect X,t)
    = \sum_{j=1}^{N_v}r_j(\vect X,t)=\sum_{j=1}^{N_v}\big(\bar{r}_j+\lambda _{j,1}r_{j,1}(\vect X,t)+\lambda _{j,2}r_{j,2}(\vect X,t)\big)
\end{equation}
where $r_j(\vect X,t)\;(j=1,2,\dots,N_v)$ denotes the $j$th component of the response $r(\vect X,t)$ induced by the $j$th non-Gaussian wind excitation $F_j(\vect X,t)$; $\bar{r}_j$ is the static component of $r_j(\vect X,t)$, obtained from a static analysis of the system subjected to $\bar{F}_j$; and $r_{j,1}(\vect X,t)$ and $r_{j,2}(\vect X,t)$ are the linear and quadratic dynamic components of $r_j(\vect X,t)$, evaluated from dynamic analyses of the system under $v_j(\vect X,t)$ and $v_j^2(\vect X,t)$, respectively.

Assuming zero initial conditions for the linear system, the explicit time-domain expressions of the response components $r_{j,1}(\vect X,t)$ and $r_{j,2}(\vect X,t)$ can be written as
\begin{equation}\label{r_j1}
\begin{array}{l}
r_{j,1}(\vect X,t_i)
    = a^{r_j}_{i,1}v_j(\vect X,t_1)+a^{r_j}_{i,2}v_j(\vect X,t_2) +\cdots+ a^{r_j}_{i,i}v_j(\vect X,t_i) \\
\hfill (i=1,2,\dots,n;j=1,2,\dots,N_v)
\end{array}
\end{equation}
\begin{equation}\label{r_j2}
\begin{array}{l}
r_{j,2}(\vect X,t_i)
    = a^{r_j}_{i,1}v_j^2(\vect X,t_1)+a^{r_j}_{i,2}v_j^2(\vect X,t_2) +\cdots+ a^{r_j}_{i,i}v_j^2(\vect X,t_i) \\
\hfill (i=1,2,\dots,n;j=1,2,\dots,N_v)
\end{array}
\end{equation}
where $n$ is the number of time steps for the dynamic analysis; $t_i=i\Delta t$ with $\Delta t$ being the time step; and $a^{r_j}_{i,1},a^{r_j}_{i,2},\dots,a^{r_j}_{i,i}\;(i=1,2,\dots,n;j=1,2,\dots,N_v)$ are the coefficients corresponding to the response $r(\vect X,t)$.

The physical meaning and computational scheme of the above coefficients were well discussed in \cite{su2014random,su2022nonstationary}. The coefficients $a^{r_j}_{i,1},a^{r_j}_{i,2},\dots,a^{r_j}_{i,i}\;(i=1,2,\dots,n)$ can be obtained via a response time-history analysis of the linear system subjected to a unit impulse excitation associated with the $j$th non-Gaussian wind excitation $F_j(\vect X,t)$. Therefore, by performing $N_v$ response time-history analyses of the linear system with any type of time-domain integration methods, the coefficients $a^{r_j}_{i,1},a^{r_j}_{i,2},\dots,a^{r_j}_{i,i}\;(i=1,2,\dots,n;j=1,2,\dots,N_v)$ with respect to any response $r(\vect X,t)$ of interest can be obtained.

Substituting Eqs.\eqref{spectral representation1}, \eqref{r_j1} and \eqref{r_j2} into Eq.\eqref{superposition}, one can obtain a compact explicit expression of $r(\vect X,t)$ as follows:
\begin{equation}\label{explicit expression}
    r(\vect X,t_i)
    = \bar{r}+\vect a_i^r \vect X+\vect X^{\mathrm T}\vect A_i^r\vect X\quad(i=1,2,\dots,n)
\end{equation}
where $\bar{r}=\sum_{j=1}^{N_v}\bar{r}_j$ is the static component of $r(\vect X,t_i)$; and $\vect a_i^r$ and $\vect A_i^r$ represent the coefficient vector and matrix corresponding to the linear and quadratic dynamic components of $r(\vect X,t_i)$, respectively, which are expressed as 
\begin{equation}\label{vect a}
    \vect a_i^r   =\sum_{j=1}^{N_v}\lambda_{j,1}\big(a^{r_j}_{i,1}\vect \psi_j(t_1)+a^{r_j}_{i,2}\vect \psi_j(t_2)+\cdots+ a^{r_j}_{i,i}\vect \psi_j(t_i) \big) \quad(i=1,2,\dots,n)
\end{equation}
\begin{equation}\label{matrix A}
    \vect A_i^r   =\sum_{j=1}^{N_v}\lambda_{j,2}\big(a^{r_j}_{i,1}\vect \psi_j^{\mathrm T}(t_1)\vect \psi_j(t_1)+a^{r_j}_{i,2}\vect \psi_j^{\mathrm T}(t_2)\vect \psi_j(t_2) +\cdots+ a^{r_j}_{i,i}\vect \psi_j^{\mathrm T}(t_i)\vect \psi_j(t_i)\big) \quad(i=1,2,\dots,n)
\end{equation}

It follows from Eq.\eqref{explicit expression} that, under non-Gaussian wind excitations, the response of the linear system at any given time instant can be expressed as a quadratic function of the input Gaussian random vector $\vect X$.

\subsection{Explicit expressions of response sensitivities}
Suppose $\theta$ is a design parameter of the linear system described in Eq.\eqref{motion equation}, and is independent of the non-Gaussian wind excitations. Then, differentiating all the terms in Eq.\eqref{motion equation} with respect to the design parameter $\theta$ leads to the following sensitivity equation as
\begin{equation}\label{sensitivity equation}
\begin{aligned}
    &\vect M\frac{\partial\ddot{\vect U}(\vect X ,t)}{\partial \theta} 
    +\vect C\frac{\partial\dot{\vect U}(\vect X ,t)}{\partial \theta} 
    + \vect K\frac{\partial\vect U(\vect X ,t)}{\partial \theta} \\
    &= \sum_{j=1}^{N_v}\frac{\partial\vect L_j}{\partial \theta} F_j(\vect X ,t) 
    -\left\{ \frac{\partial\vect M}{\partial \theta} \ddot{\vect U}(\vect X ,t)
    +\frac{\partial\vect C}{\partial \theta} \dot{\vect U}(\vect X ,t)
    + \frac{\partial\vect K}{\partial \theta} \vect U(\vect X ,t) \right\}
\end{aligned}
\end{equation}

Differentiation of the both sides of Eq.\eqref{superposition} with respect to the design parameter $\theta$ yields the response sensitivity $\partial r(\vect X,t)/\partial \theta$ as follows:
\begin{equation}\label{superposition sensitivity}
    \frac{\partial r(\vect X,t)}{\partial \theta}
    = \sum_{j=1}^{N_v}\frac{\partial r_j(\vect X,t)}{\partial \theta}=\sum_{j=1}^{N_v}\left(\frac{\partial \bar{r}_j}{\partial \theta }+\lambda _{j,1} \frac{\partial r_{j,1}(\vect X,t)}{\partial \theta}+\lambda _{j,2}\frac{\partial r_{j,2}(\vect X,t)}{\partial \theta}\right)
\end{equation}
where the sensitivity of static component $\partial \bar{r}_j/\partial \theta$ can be obtained from a static sensitivity analysis of the system under $\bar{F}_j$; and the sensitivities of linear and quadratic components, $\partial r_{j,1}(\vect X,t)/\partial \theta$ and $\partial r_{j,2}(\vect X,t)/\partial \theta$, need to be evaluated from dynamic sensitivity analyses of the system subjected to $v_j(\vect X,t)$ and $v_j^2(\vect X,t)$, respectively.

Differentiating Eqs.\eqref{r_j1} and \eqref{r_j2} with respect to the design parameter $\theta$, one can obtain the explicit time-domain expressions of the sensitivities of response components, $\partial r_{j,1}(\vect X,t)/\partial \theta$ and $\partial r_{j,2}(\vect X,t)/\partial \theta$, as follows:
\begin{equation}\label{r_j1_sen}
\begin{array}{l}
\dfrac{\partial r_{j,1}(\vect X,t_i)}{\partial \theta}
    = b^{r_j}_{i,1}v_j(\vect X,t_1)+b^{r_j}_{i,2}v_j(\vect X,t_2) +\cdots+ b^{r_j}_{i,i}v_j(\vect X,t_i) \\
\hfill (i=1,2,\dots,n;j=1,2,\dots,N_v)
\end{array}
\end{equation}
\begin{equation}\label{r_j2_sen}
\begin{array}{l}
\dfrac{\partial r_{j,2}(\vect X,t_i)}{\partial \theta}
    = b^{r_j}_{i,1}v_j^2(\vect X,t_1)+b^{r_j}_{i,2}v_j^2(\vect X,t_2) +\cdots+ b^{r_j}_{i,i}v_j^2(\vect X,t_i) \\
\hfill (i=1,2,\dots,n;j=1,2,\dots,N_v)
\end{array}
\end{equation}
where $b^{r_j}_{i,1}=\partial a^{r_j}_{i,1}/\partial \theta,b^{r_j}_{i,2}=\partial a^{r_j}_{i,2}/\partial \theta,\dots,b^{r_j}_{i,i}=\partial a^{r_j}_{i,i}/\partial \theta\;(i=1,2,\dots,n;j=1,2,\dots,N_v)$ are the coefficients corresponding to the response sensitivity $\partial r(\vect X,t)/\partial \theta$.

Similar to the coefficients $a^{r_j}_{i,1},a^{r_j}_{i,2},\dots,a^{r_j}_{i,i}\;(i=1,2,\dots,n;j=1,2,\dots,N_v)$ shown in Eqs.\eqref{r_j1} and \eqref{r_j2}, the coefficients $b^{r_j}_{i,1},b^{r_j}_{i,2},\dots,b^{r_j}_{i,i}\;(i=1,2,\dots,n;j=1,2,\dots,N_v)$ can also be computed based on their inherent physical meanings, and a computational scheme was provided in \cite{hu2016explicit,xian2023reliability}. The coefficients $b^{r_j}_{i,1},b^{r_j}_{i,2},\dots,b^{r_j}_{i,i}\;(i=1,2,\dots,n)$ can be obtained through a response sensitivity time-history analysis of the linear system exposed to a unit impulse excitation associated with the $j$th non-Gaussian wind excitation $F_j(\vect X,t)$. Therefore, the coefficients $b^{r_j}_{i,1},b^{r_j}_{i,2},\dots,b^{r_j}_{i,i}\;(i=1,2,\dots,n;j=1,2,\dots,N_v)$ for the sensitivity of any response $\partial r(\vect X,t)/\partial \theta$ can be obtained by conducting $N_v$ response sensitivity time-history analyses of the linear system. Such analyses can be achieved using any type of time-domain integration methods.

Following a procedure similar to that used in deriving Eq.\eqref{explicit expression}, substitution of Eqs.\eqref{spectral representation1}, \eqref{r_j1_sen} and \eqref{r_j2_sen} into Eq.\eqref{superposition sensitivity} yields a compact explicit expression of $\partial r(\vect X,t)/\partial \theta$ as follows:
\begin{equation}\label{explicit expression_sen}
    \frac{\partial r(\vect X,t_i)}{\partial \theta}
    = \frac{\partial \bar{r}}{\partial \theta}+\vect b_i^r \vect X+\vect X^{\mathrm T}\vect B_i^r\vect X\quad(i=1,2,\dots,n)
\end{equation}
where
$\partial \bar{r}/\partial \theta=\sum_{j=1}^{N_v}\partial \bar{r}_j/\partial \theta$ denotes the sensitivity of the static component of $r(\vect X,t_i)$; and $\vect b_i^r$ and $\vect B_i^r$ denote the coefficient vector and matrix corresponding to the sensitivities of the linear and quadratic dynamic components of $r(\vect X,t_i)$, respectively, which are expressed as
\begin{equation}\label{vect b}
    \vect b_i^r   =\sum_{j=1}^{N_v}\lambda_{j,1}\big(b^{r_j}_{i,1}\vect \psi_j(t_1)+b^{r_j}_{i,2}\vect \psi_j(t_2)+\cdots+ b^{r_j}_{i,i}\vect \psi_j(t_i) \big) \quad(i=1,2,\dots,n)
\end{equation}
\begin{equation}\label{matrix B}
    \vect B_i^r   =\sum_{j=1}^{N_v}\lambda_{j,2}\big(b^{r_j}_{i,1}\vect \psi_j^{\mathrm T}(t_1)\vect \psi_j(t_1)+b^{r_j}_{i,2}\vect \psi_j^{\mathrm T}(t_2)\vect \psi_j(t_2) +\cdots+ b^{r_j}_{i,i}\vect \psi_j^{\mathrm T}(t_i)\vect \psi_j(t_i)\big) \quad(i=1,2,\dots,n)
\end{equation}

Thus far, the explicit expressions of both the dynamic responses and their sensitivities of the linear system under non-Gaussian wind excitations have been established, as given in Eqs.\eqref{explicit expression} and \eqref{explicit expression_sen}, respectively.
These concise closed-form expressions greatly facilitates the first-passage dynamic reliability analysis and the corresponding reliability sensitivity analysis, which will be elaborated in Sections \ref{sec:first-passage dynamic reliability} and \ref{sec:Sensitivity analysis of first-passage dynamic reliability}.

\section{First-passage dynamic reliability analysis}\label{sec:first-passage dynamic reliability}
\subsection{Component dynamic reliability}
For dynamic reliability problems of linear systems, a component failure event can be defined as the exceedance of the prescribed threshold by a response component of interest at a given time instant. Suppose that $m$  critical response components and $n$ discrete time steps are considered. This results in a total of $mn$ component failure events, which can be expressed as
\begin{equation}\label{component failure event}
    F_{ki}=\left\{ c_k -  s_{k,i}(\vect X)  \leq0 \right\}\quad(i=1,2,\dots,n;k=1,2,\dots,m)
\end{equation}
where $s_{k,i}(\vect X)=s_{k}(\vect X,t_i)$ is the $k$th critical response component $s_k(\vect X,t)$ at the time instant $t_i$; and $c_k$ is the positive threshold for the $k$th critical response component $s_k(\vect X,t)$. Without loss of generality, only up-crossing of positive thresholds is considered to simplify the notation. The extension to down-crossing of negative thresholds or simultaneous consideration of both cases is straightforward.

The limit-state function with respect to the component failure event $F_{ki}$ is written as
\begin{equation}\label{component lsf}
    g_{ki}(\vect x) = c_k - s_{k,i}(\vect x)\quad(i=1,2,\dots,n;k=1,2,\dots,m)
\end{equation}
Let $r(\vect X,t)$ be $s_k(\vect X,t)$. Substituting Eq.\eqref{explicit expression} into Eq.\eqref{component lsf}, one yields
\begin{equation}\label{component lsf1}
    g_{ki}(\vect x) = c_k - (\bar{s}_k+\vect a_i^{s_k} \vect x+\vect x^{\mathrm T}\vect A_i^{s_k}\vect x)\quad(i=1,2,\dots,n;k=1,2,\dots,m)
\end{equation}

\begin{figure}[ht]
  \centering
  \includegraphics[width=0.5\textwidth] {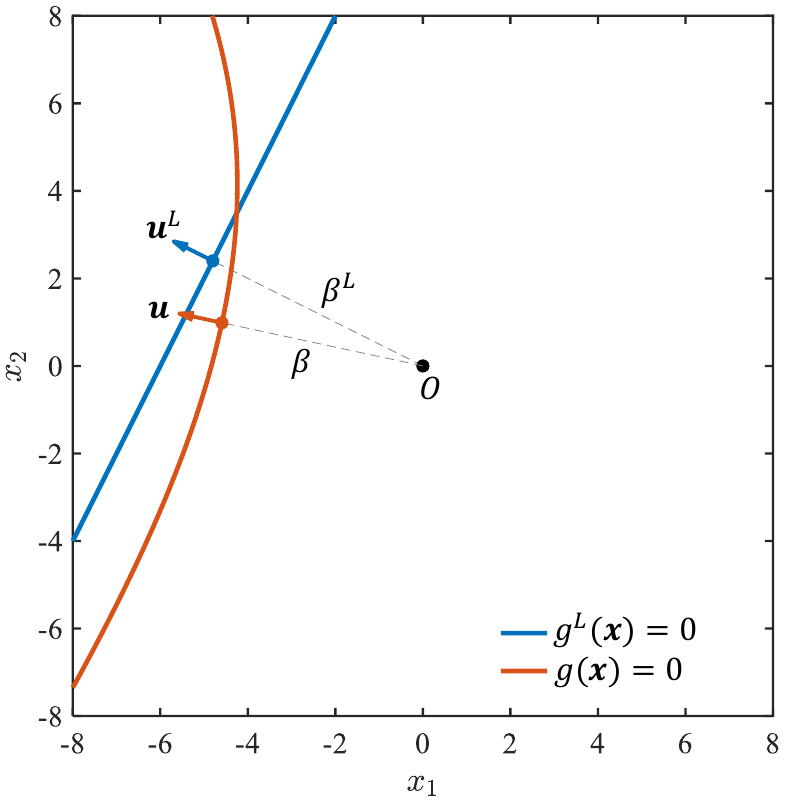}
  \caption{\color{Black}\textbf{Geometrical illustration of quadratic component limit state.} Consider a quadratic component limit-state function $g(\vect x)=c-\vect a \vect x-\vect x^{\mathrm T}\vect A\vect x$, where $\vect x=[x_1\;x_2]^{\mathrm T}$, $c=12$, $\vect a=[-2,1]$, $\vect A=
\left[\begin{smallmatrix}
0.05 & 0.02\\
0.02 & -0.10
\end{smallmatrix}\right]$. Omitting the quadratic term $\vect x^{\mathrm T}\vect A\vect x$ yields the linear limit-state function $g^L(\vect x)=c-\vect a \vect x$. The primary branch of the hyperbolic failure boundary $g(\vect x)=0$ can be regarded as a slightly distorted version of the linear failure boundary $g^L(\vect x)=0$, whereas the secondary branch lies outside the region of interest and therefore contributes negligibly to the failure probability. }
  \label{fig:1}
\end{figure}

Since the coefficient matrix $\mathbf A_i^{s_k}$ is indefinite, as evidenced in Eq.\eqref{matrix A}, the limit-state hypersurface $S_{ki}=\left \{ \vect x \in \mathbb{R}^d:g_{ki}(\vect x)=0\right\}$ is a hyperbolic-type quadratic hypersurface. For the present wind-induced dynamic reliability problem, the quadratic component $\vect x^{\mathrm T}\vect A_i^{s_k}\vect x$ contributes much less to the structural response than the linear component $\vect a_i^{s_k} \vect x$. Consequently, one branch of the quadratic hypersurface can be interpreted as a slight perturbation of the linear hyperplane $g^L_{ki}(\vect x) = c_k - (\bar{s}_k+\vect a_i^{s_k} \vect x)=0$. The existence of an additional branch is a mathematical consequence of the quadratic representation. Nevertheless, this branch is located much farther from the origin than the primary branch, and thus its contribution to the failure probability can be regarded as negligible. As the quadratic contribution tends to zero, the primary branch of the quadratic hypersurface continuously degenerates into the linear hyperplane, while the secondary branch recedes toward infinity. A geometrical illustration of the quadratic component limit state is presented in Figure \ref{fig:1}.

The probability of the component failure event $F_{ki}$ is given by
\begin{equation}\label{component failure probability}
    \mathbb{P}_{ki}
    = \int_{\mathbb{R}^d} \mathcal{H}(-g_{ki}(\vect x)) f_{\vect X}(\vect x)\, d\vect x\quad(i=1,2,\dots,n;k=1,2,\dots,m)
\end{equation}
where $f_{\vect X}(\vect x)$ is the joint probability density function (PDF) of the standard Gaussian random vector $\vect X$; and $\mathcal{H}(\cdot)$ denotes the Heaviside step function, with $\mathcal{H}(-g_{ki}(\vect x))$  taking 1 if $g_{ki}(\vect x)\leq 0$ and 0 otherwise.

The design point is defined as the point on the quadratic limit-state hypersurface $S_{ki}$ that has the smallest distance from the origin, which can be obtained by solving the following constrained optimization problem:
\begin{equation}\label{optimization}
\vect{x}_{ki}^{\star}
=
\arg\min_{\vect{x}}
\frac12\vect{x}^{\mathrm T}\vect{x},
\quad
\text{s.t. }
g_{ki}(\vect{x})=0\quad(i=1,2,\dots,n;k=1,2,\dots,m)
\end{equation}
The above optimization problem is solved using the gradient projection method \cite{Rosen1961gradient}, where the gradient of the component limit-state function can be derived from Eq.\eqref{component lsf1} as follows:
\begin{equation}\label{gradient}
    \nabla _{\vect x}g_{ki}(\vect{x})= -(\vect a_i^{s_k})^{\mathrm T}-2\vect A_i^{s_k}\vect x\quad(i=1,2,\dots,n;k=1,2,\dots,m)
\end{equation}

Once the design point is obtained, the corresponding unit vector pointing from the origin to the design point and the associated reliability index, defined as the Euclidean distance from the origin to the design point, are given by
\begin{equation}\label{unit vector}
    \vect{u}_{ki} = \frac{\vect{x}_{ki}^{\star}}{\left\| \vect{x}_{ki}^{\star} \right\|}\quad(i=1,2,\dots,n;k=1,2,\dots,m)
\end{equation}
\begin{equation}\label{beta}
    \beta_{ki} = \left\| \vect{x}_{ki}^{\star} \right\|\quad(i=1,2,\dots,n;k=1,2,\dots,m)
\end{equation}
Then, the first-order approximation of the component failure probability is obtained as
\begin{equation}\label{component probability approximate}
    \tilde{\mathbb{P}}_{ki} = \Phi (-\beta_{ki})\quad(i=1,2,\dots,n;k=1,2,\dots,m)
\end{equation}
where $\Phi (\cdot)$ denotes the cumulative distribution function (CDF) of the standard Gaussian distribution. 

An accurate estimate of the component failure probability $\mathbb{P}_{ki}$ in Eq.\eqref{component failure probability} can be obtained by line sampling, in which the sampling direction vector is chosen as the unit vector $\vect{u}_{ki}$. It should be noted that, although direct estimation of the component failure probability is necessary for the subsequent first-passage dynamic reliability analysis \cite{Wang2016efficient}, it is not required for the sensitivity analysis of first-passage dynamic reliability, where only its first-order approximation as presented in Eq.\eqref{component probability approximate} is sufficient. The rationale behind this will be further discussed in Section \ref{sec:Sensitivity analysis of first-passage dynamic reliability}.

\subsection{First-passage dynamic reliability}
 For first-passage dynamic reliability problems of linear systems, the system failure event can be formulated as the union of the component failure events, i.e.
\begin{equation}\label{system failure event}
F=\bigcup_{k=1}^{m}\bigcup_{i=1}^{n}F_{ki}
\end{equation}
and the corresponding system limit-state function is expressed as
\begin{equation}\label{system lsf}
    G(\vect x)= \min_{k=1}^{m}\min_{i=1}^{n}g_{ki}(\vect x)
\end{equation}
Then, the probability of the system failure event $F$ can be obtained as
\begin{equation}\label{system failure probability}
    \mathbb{P}
    = \int_{\mathbb{R}^d} \mathcal{H}(-G(\vect x)) f_{\vect X}(\vect x)\, d\vect x
\end{equation}

Under Gaussian excitations, where the component limit-state function $g_{ki}(\vect x)$ is linear rather than quadratic, several advanced simulation techniques, such as the importance sampling with elementary events \cite{au2001first} and the domain decomposition method \cite{katafygiotis2006domain}, have been proposed for first-passage dynamic reliability analysis of linear systems at high efficiency. Among these approaches, the domain decomposition method has been further extended to handle the non-Gaussian wind excitations. Such an extension requires the use of line sampling to address two technical issues, i.e., the computation of component failure probabilities and sampling from the component failure domains \cite{Wang2016efficient}.

\section{Sensitivity analysis of first-passage dynamic reliability }\label{sec:Sensitivity analysis of first-passage dynamic reliability}
\subsection{Surface decomposition method}
Suppose that $\theta$ is a design parameter of the structure. Then, differentiating the system failure probability shown in Eq.\eqref{system failure probability} with respect to $\theta$, one obtains
\begin{equation}\label{system failure probability sensitivity}
    \frac{\partial\mathbb{P}}{\partial \theta}
    = -\int_{\mathbb{R}^d} \delta(-G(\vect x)) \frac{\partial G(\vect x)}{\partial \theta}f_{\vect X}(\vect x)\, d\vect x
\end{equation}
where $\delta(\cdot)$ denotes the Dirac delta function; and $\partial G(\vect x)/\partial \theta$ denotes the sensitivity of the system limit-state function $G(\vect x)$ with respect to $\theta$.

Leveraging the property of Dirac delta function, the $d$-dimensional volume integral in Eq.\eqref{system failure probability sensitivity} can be converted into the following $(d-1)$-dimensional surface integral \cite{hormander2003analysis}:
\begin{equation}\label{system surface integral}
    \frac{\partial\mathbb{P}}{\partial \theta}
    = -\int_{S} \frac{1}{\left\|\nabla _{\vect x}G(\vect x)\right\|} \frac{\partial G(\vect x)}{\partial \theta}f_{\vect X}(\vect x)\, d S
\end{equation}
where $d S$ denotes the differential element of the $(d-1)$-dimensional system limit-state hypersurface $S=\left \{ \vect x \in \mathbb{R}^d:G(\vect x)=0\right\}$; and $\left\|\nabla _{\vect x}G(\vect x)\right\|$ denotes the Euclidean norm of the gradient of the system limit-state function $G(\vect x)$. 

As the system limit-state hypersurface $S$ is highly non-smooth and high-dimensional, direct evaluation of the system surface integral shown in Eq.\eqref{system surface integral} is computationally challenging. The surface decomposition method \cite{xian2026surface} addresses this challenge by decomposing $S$ into a collection of truncated smooth component limit-state hypersurfaces $S^\mathrm{t}_{ki}$, defined as
\begin{equation}\label{component hypersurface}
S^\mathrm{t}_{ki}=\left \{ \vect x \in \mathbb{R}^d:g_{ki}(\vect x)=0,\;\mathbb{I}_{ki}(\vect x)=1\right\}\quad(i=1,2,\dots,n;k=1,2,\dots,m)
\end{equation}
where $\mathbb{I}_{ki}(\vect x)$ is the indicator function defined as
\begin{equation}\label{indicator}
\mathbb{I}_{ki}(\vect x) =
\begin{cases}
1, & \text{if } g_{ki}(\vect x) = \min\limits_{j=1}^{m}\min\limits_{s=1}^{n} g_{js}(\vect x), \\[6pt]
0, & \text{otherwise}.
\end{cases}\quad(i=1,2,\dots,n;k=1,2,\dots,m)
\end{equation}
This indicator function serves to identify the portion of each component limit-state hypersurface $S_{ki}$ that actually defines the system limit-state hypersurface $S$, so that the collection of all $S^\mathrm{t}_{ki}$ forms a non-overlapping decomposition of $S$.

\begin{figure}[ht]
  \centering
  \includegraphics[width=1\textwidth] {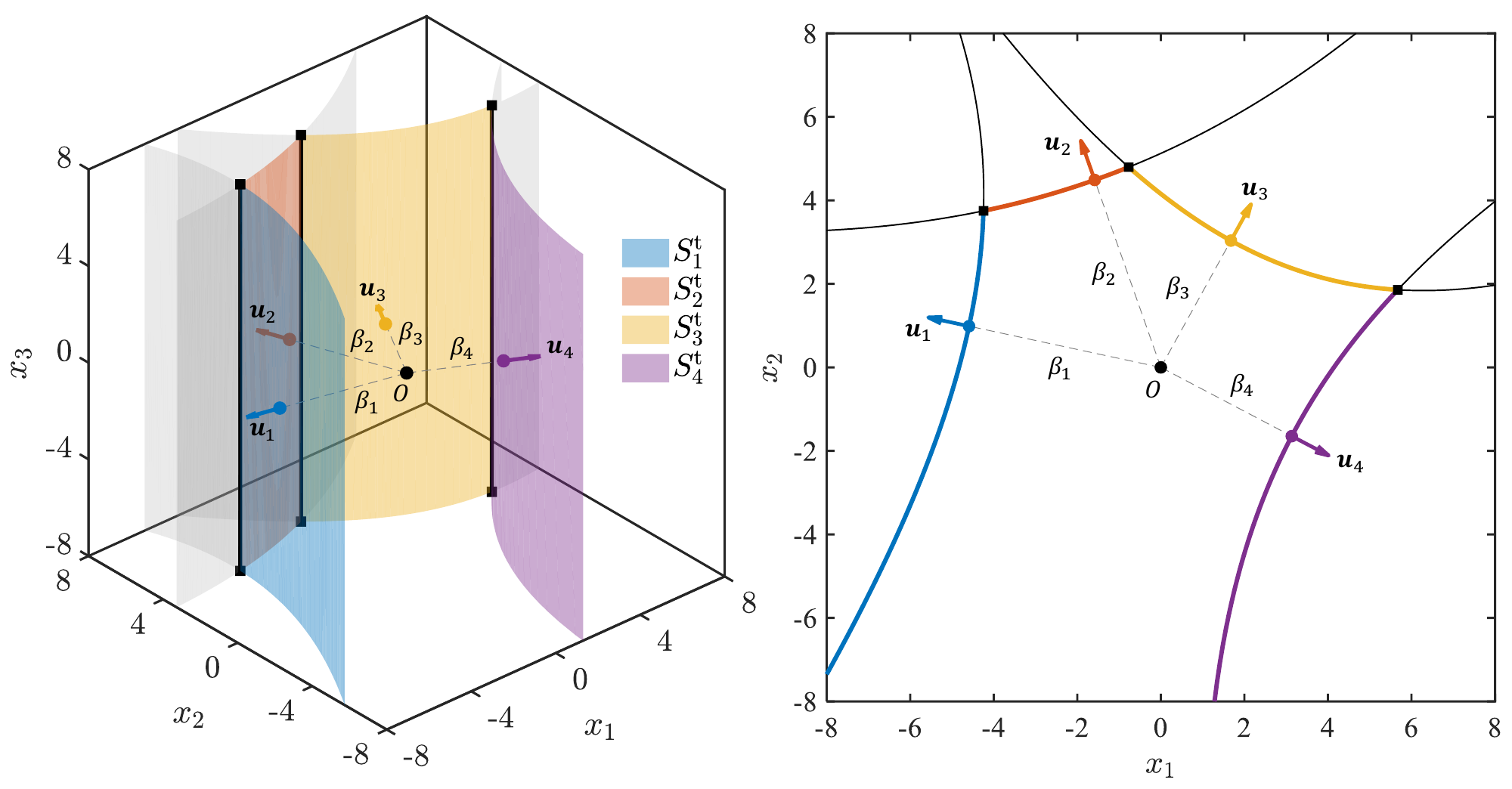}
  \caption{\color{Black}\textbf{Geometrical illustration of surface decomposition method under quadratic Gaussian case.} Consider a series system limit-state function $G(\vect x)=\min_{k=1}^{4} g_k(\vect x)$. The four component limit-state functions are $g_k(\vect x)=c_k-\vect a_k\vect x-\vect x^{\mathrm T} \vect A_k\vect x \;(k=1,2,3,4)$, where $\vect x=[x_1\;x_2]^{\mathrm T}$, $c_1=12,\;c_2=18,\;c_3=36,\;c_4=10$, $\vect a_1=[-2,1]$, $\vect a_2=[-1,3]$, $\vect a_3=[6,7]$, $\vect a_4=[2,-1]$, $\vect A_1=\left[\begin{smallmatrix}0.05 & 0.02\\0.02 & -0.10\end{smallmatrix}\right]$, $\vect A_2=\left[\begin{smallmatrix}-0.08 & -0.08\\-0.08 & 0.10\end{smallmatrix}\right]$, $\vect A_3=\left[\begin{smallmatrix}-0.55 & 0.25\\0.25 & 0.40\end{smallmatrix}\right]$, $\vect A_4=\left[\begin{smallmatrix}0.10 & -0.12\\-0.12 & -0.05\end{smallmatrix}\right]$. The above limit-state functions are extended to three-dimensional space by introducing an additional variable $x_3$ but with zero coefficients. The system limit-state surface $S=\left \{ \vect x \in \mathbb{R}^3:G(\vect x)=0\right\}$ can be decomposed into four truncated component limit-state surfaces $S^\mathrm{t}_k=\left \{ \vect x \in \mathbb{R}^3:g_{k}(\vect x)=0,\;\mathbb{I}_{k}(\vect x)=1\right\}\;(k=1,2,3,4)$, where the indicator function $\mathbb{I}_k(\vect x)$ takes 1 if $g_k(\vect x)=\min_{j=1}^{4} g_{j}(\vect x)$, and 0 otherwise. The sensitivity of series system failure probability can be expressed as the sum of four surface integrals over $S_k^{\mathrm{t}}\;(k=1,2,3,4)$. }
\label{fig:2}
\end{figure}

Accordingly, the system surface integral in Eq.\eqref{system surface integral} can be decomposed as
\begin{equation}\label{system failure probability sensitivity expanded}
    \frac{\partial\mathbb{P}}{\partial \theta}
    = \sum_{k=1}^{m}\sum_{i=1}^{n}\eta_{ki} 
\end{equation}
where 
\begin{equation}\label{component surface integral}
\begin{split}
    \eta_{ki}
    &= -\int_{S^\mathrm{t}_{ki}} \frac{1}{\left\|\nabla _{\vect x}g_{ki}(\vect x)\right\|} \frac{\partial g_{ki}(\vect x)}{\partial \theta}f_{\vect X}(\vect x)\, d S^\mathrm{t}_{ki}
    \\&=-\int_{S_{ki}} \frac{1}{\left\|\nabla _{\vect x}g_{ki}(\vect x)\right\|} \frac{\partial g_{ki}(\vect x)}{\partial \theta}\mathbb{I}_{ki}(\vect x)f_{\vect X}(\vect x)\, d S_{ki}\quad(i=1,2,\dots,n;k=1,2,\dots,m)
    \end{split}
\end{equation}
denotes the component surface integral associated with the truncated hypersurface $S^\mathrm{t}_{ki}$, on which the system limit-state function satisfies $G(\vect x)=g_{ki}(\vect x)$.
A geometrical illustration of the surface decomposition method under quadratic Gaussian case is presented in Figure \ref{fig:2}.

In general, not all component surface integrals need to be evaluated, since some make negligible contributions to the system failure probability sensitivity $\partial\mathbb{P}/{\partial \theta}$. To improve the computational efficiency, it is therefore desirable to identify the dominant component surface integrals. To this end, an importance sampling strategy is adopted to estimate the summation in Eq.\eqref{system failure probability sensitivity expanded}. Specifically, Eq.\eqref{system failure probability sensitivity expanded} can be rewritten as
\begin{equation}\label{system failure probability sensitivity expanded rewritten}
    \frac{\partial\mathbb{P}}{\partial \theta}
    = mn\sum_{k=1}^{m}\sum_{i=1}^{n}\eta_{ki} f(k,i)
\end{equation}
where $f(k,i)=1/(mn)$ is the discrete uniform probability mass function (PMF) defined over the index pairs $(k,i)\;(i=1,2,\dots,n;k=1,2,\dots,m)$.

Introducing an importance sampling PMF $h(k,i)$ to Eq.\eqref{system failure probability sensitivity expanded rewritten} yields 
\begin{equation}\label{system failure probability sensitivity IS}
    \frac{\partial\mathbb{P}}{\partial \theta}
    = mn\sum_{k=1}^{m}\sum_{i=1}^{n}\eta_{ki} \frac{f(k,i)}{h(k,i)}h(k,i)=\sum_{k=1}^{m}\sum_{i=1}^{n}\eta_{ki} \frac{1}{h(k,i)}h(k,i)
\end{equation}
When the exact component failure probabilities are analytically available, as is the case for linear systems under Gaussian excitations \cite{xian2026surface}, one can construct
the importance sampling PMF based on the relative magnitudes of these probabilities, i.e. 
\begin{equation}\label{constructed ID}
    h(k,i)
    = \frac{\mathbb{P}_{ki}}{ \sum_{j=1}^{m}\sum_{s=1}^{n}\mathbb{P}_{js} }
\end{equation}
The rationale behind this choice is that a higher component failure probability $\mathbb{P}_{ki}$ generally indicates a larger surface measure of the truncated hypersurface $S^\mathrm{t}_{ki}$. Consequently, the associated surface integral $\eta_{ki}$ contributes more significantly to the system failure probability sensitivity.

However, for quadratic Gaussian case, the component failure probabilities cannot be derived analytically and thus need to be evaluated using line sampling. In fact, the exact values of the component failure probabilities are not required, while only their relative magnitudes matter. As the relative magnitudes of the exact probabilities are close to those of their first‑order approximations, one can also construct the importance sampling PMF based on the relative magnitudes of these approximations, i.e.
\begin{equation}\label{constructed ID approximate}
    h(k,i)
    = \frac{\tilde{\mathbb{P}}_{ki}}{ \sum_{j=1}^{m}\sum_{s=1}^{n}\tilde{\mathbb{P}}_{js} }
\end{equation}
where $\tilde{\mathbb{P}}_{ki}$ is the first-order approximation of $\mathbb{P}_{ki}$, which has been obtained in Eq.\eqref{component probability approximate}.

\subsection{Nested sampling algorithm}
The component surface integral $\eta_{ki}$ in Eq.\eqref{component surface integral} is evaluated using line sampling, which projects the limit-state hypersurface onto an auxiliary hyperplane and performs one-dimensional searches along prescribed sampling directions. Accordingly, $\eta_{ki}$ can be rewritten as
\begin{equation}\label{component surface integral LS}
    \eta_{ki}
    = \displaystyle\int_{\mathbb{R}^d} z_{ki}(\vect x)f_{\vect X}(\vect x)\, d \vect x \quad(i=1,2,\dots,n;k=1,2,\dots,m)
\end{equation}
\begin{equation}\label{z}
    z_{ki}(\vect x) =-\dfrac{1}{\left|\vect u_{ki}^{\mathrm T}\nabla _{\vect x}g_{ki}(\vect x_{ki})\right|} \dfrac{\partial g_{ki}(\vect x_{ki})}{\partial \theta}\mathbb{I}_{ki}(\vect x_{ki})\varphi (p_{ki})\quad(i=1,2,\dots,n;k=1,2,\dots,m)
\end{equation}
where $\vect x_{ki}=\vect x+\Delta p_{ki}\vect u_{ki}$ denotes the intersection point of the sampling line and the hypersurface $S_{ki}$; $\Delta p_{ki}$ is the signed distance along the sampling direction $\vect u_{ki}$ from $\vect x$ to $\vect x_{ki}$, which can be determined by solving 
$g_{ki}(\vect x+\Delta p_{ki}\vect u_{ki})=0$;  $p_{ki}=\vect u_{ki}^{\mathrm T}\vect x+\Delta p_{ki}$ is the coordinate of $\vect x_{ki}$ along the sampling direction $\vect u_{ki}$; and $\varphi (\cdot)$ is the standard Gaussian PDF. The derivation of Eqs.\eqref{component surface integral LS} and \eqref{z} using line sampling is presented in \ref{appendix a}.

From Eq.\eqref{component lsf}, the sensitivity of the component limit-state function $g_{ki}(\vect x)$ with respect to $\theta$ can be derived as
\begin{equation}\label{sensitivity of component lsf}
    \frac{\partial g_{ki}(\vect x)}{\partial \theta} = - \frac{\partial s_{k,i}(\vect x)}{\partial \theta}\quad(i=1,2,\dots,n;k=1,2,\dots,m)
\end{equation}
where $\partial s_{k,i}(\vect x)/\partial \theta$ is the sensitivity of the $k$th critical response component $s_k(\vect X,t)$ at the time instant $t_i$ with respect to $\theta$. Let $\partial r(\vect X,t)/\partial \theta = \partial s_k(\vect X,t)/\partial \theta$. Substituting Eq.\eqref{explicit expression_sen} into Eq.\eqref{sensitivity of component lsf}, one obtains
\begin{equation}\label{sensitivity of component lsf1}
    \frac{\partial g_{ki}(\vect x)}{\partial \theta} = - \left (\frac{\partial \bar{s}_k}{\partial \theta}+\vect b_i^{s_k} \vect x+\vect x^{\mathrm T}\vect B_i^{s_k}\vect x \right )\quad(i=1,2,\dots,n;k=1,2,\dots,m)
\end{equation}

From Eq.\eqref{component lsf1}, solving $g_{ki}(\vect x+\Delta p_{ki}\vect u_{ki})=0$ reduces to solving the following quadratic equation for $\Delta p_{ki}$:
\begin{equation}\label{quadratic equation}
    -\vect{u}_{ki}^{\mathrm T}\vect A_i^{s_k}\vect{u}_{ki}(\Delta p_{ki})^2-\vect{u}_{ki}^{\mathrm T} \nabla _{\vect x}g_{ki}(\vect x)\Delta p_{ki}+g_{ki}(\vect x)=0\quad (i=1,2,\dots,n;k=1,2,\dots,m)
\end{equation}
For the problems investigated in this study, the above quadratic equation admits two real roots. The root with the smaller absolute value is selected, as it corresponds to the closest intersection point between the sampling line and the component limit-state hypersurface. The other root is a large negative value, which corresponds to a distant intersection point on the opposite branch of the hyperbolic limit-state hypersurface and is therefore discarded.

According to Eq.\eqref{indicator}, evaluating the indicator function $\mathbb{I}_{ki}(\vect x_{ki})$ in Eq.\eqref{z} requires the values of all component limit-state functions at $\vect x_{ki}$. However, from a geometrical perspective, $\mathbb{I}_{ki}(\vect x_{ki})$ can be evaluated without explicitly computing these limit-state functions. Since $g_{ki}(\vect x_{ki})=0$, $\mathbb{I}_{ki}(\vect x_{ki})$ simply determines whether $\vect x_{ki}$ lies in the mutual safe domain of all the other component failure limit-state functions. Geometrically, this is equivalent to checking whether, starting from the same sampling point $\vect x$, the projection of the increment $\Delta p_{ki}\vect u_{ki}$ onto every other sampling direction is smaller than the corresponding increment along that direction, i.e.
\begin{equation}\label{indicator1}
\begin{aligned}
\mathbb{I}_{ki}(\vect x_{ki}) &=
\begin{cases}
1, & \text{if } \Delta p_{ki} \vect u_{ki}^{\mathrm T} \vect u_{js}<\Delta p_{js},\;\forall (j,s)\neq(k,i), \\[6pt]
0, & \text{otherwise}.
\end{cases} \quad (i=1,2,\dots,n; k=1,2,\dots,m)
\end{aligned}
\end{equation}

Substituting Eqs.\eqref{component surface integral LS} and \eqref{z} into Eq.\eqref{system failure probability sensitivity IS} yields 
\begin{equation}\label{EX}
    \frac{\partial\mathbb{P}}{\partial \theta}
    =\mathbb E_{\vect X}[y(\vect X)]=\displaystyle\int_{\mathbb{R}^d} y(\vect x)f_{\vect X}(\vect x)\, d \vect x 
\end{equation}
where
\begin{equation}\label{y(x)}
    y(\vect x) = \mathbb E_h \!\left[ \frac{z_{KI}(\vect x)}{h(K,I)} \right]=\sum_{k=1}^{m}\sum_{i=1}^{n}\dfrac{z_{ki}(\vect x)}{h(k,i)}h(k,i) 
\end{equation}
Then, the system failure probability sensitivity $\partial\mathbb{P}/\partial \theta$ can be evaluated using the nested Monte Carlo estimator as follows:
\begin{equation}\label{estimator1}
    \widehat{\frac{\partial\mathbb P}{\partial\theta}}
    = \frac{1}{N_{\mathrm {out}}}\sum_{q=1}^{N_{\mathrm {out}}}\hat y(\vect x_{q})
\end{equation}
\begin{equation}\label{estimator2}
    \hat y(\vect x_{q})
   =\frac{1}{N_{\mathrm {in}}}\sum_{j=1}^{N_{\mathrm {in}}}\frac{z_{k_ji_j}(\vect x_{q})}{h(k_j,i_j)}
\end{equation}
where $\vect x_{q}\;(q=1,2,\dots,N_{\mathrm {out}})$ are independent samples drawn from the PDF $f_{\vect X}(\vect x)$; $N_{\mathrm {out}}$ denotes the number of outer-level samples; $(k_j,i_j)\;(j=1,2,\dots,N_{\mathrm {in}})$ are independent index pairs drawn from the importance sampling PMF $h(k,i)$ defined in Eq.\eqref{constructed ID approximate}; and $N_{\mathrm {in}}$ denotes the number of inner-level samples.

In practice, the number of inner-level samples can be fixed at a prescribed value, e.g., $N_{\mathrm {in}}=100$, whereas the number of outer-level samples is increased iteratively until the coefficient of variation (COV) of the estimator falls below a prescribed tolerance, e.g., $tol=0.1$. Note that, in the proposed nested sampling algorithm, the number of evaluations of the system limit-state function $G(\vect x)$ is equal to the number of outer-level samples, $N_{\mathrm {out}}$, and is independent of the number of inner-level samples, $N_{\mathrm {in}}$. Specifically, for a given outer-level sample $\vect x_q$, evaluating $\hat{y}(\vect x_q)$ in Eq.\eqref{estimator2} requires the values of the indicator functions in Eq.\eqref{indicator1} for all sampled index pairs $(k_j,i_j)\;(j=1,2,\dots,N_{\mathrm {in}})$. To this end, the corresponding scalars $\Delta p_{ki}\;(i=1,2,\dots,n;k=1,2,\dots,m)$ must first be determined for all component limit-state functions. This means that, according to Eq.\eqref{quadratic equation}, all component limit-state functions at $\vect x_q$, i.e., $g_{ki}(\vect x_q)\;(i=1,2,\dots,n;k=1,2,\dots,m)$ need to be computed, which is equivalent to a single evaluation of the system limit-state function at $\vect x_q$, i.e., $G(\vect x_q)$.

The nested Monte Carlo estimator presented in Eqs.\eqref{estimator1} and \eqref{estimator2} is unbiased. Its mean and variance can be derived as
\begin{equation}\label{estimator mean}
\mathbb E \!\left[\widehat{\frac{\partial\mathbb{P}}{\partial\theta}}\right]=\mathbb E_{\vect X}\!\left[y(\vect X)\right]
\end{equation}
\begin{equation}\label{estimator variance}
\mathbb Var \!\left[\widehat{\frac{\partial\mathbb{P}}{\partial\theta}}\right]=\frac{1}{N_{\mathrm{out}}}\left[\mathbb Var_{\vect X}\!\left[y(\vect X)\right]+\mathbb{E}_{\vect X}\!
\left[\frac{1}{N_{\mathrm{in}}}\mathbb Var_{h}\!\left(\frac{z_{KI}(\vect X)}{h(K,I)}\middle|\vect X\right)\right]\right]
\end{equation}
The derivation of Eqs.\eqref{estimator mean} and \eqref{estimator variance} is provided in \ref{appendix b}. 

The estimator itself shown in Eq.\eqref{estimator1} serves as the sample estimate of its mean. The sample estimate of the variance of the estimator is given by
\begin{equation}\label{estimator variance approx}
\widehat{\mathbb Var }\!\left[\widehat{\frac{\partial\mathbb{P}}{\partial\theta}}\right]=\frac{1}{N_{\mathrm{out}}}(\widehat{\mathbb Var }_{\mathrm {out}} + \widehat{\mathbb Var }_{\mathrm {in}})
\end{equation}
where $\widehat{\mathbb Var }_{\mathrm {out}}$ is the outer-level sampling variance, i.e.
\begin{equation}\label{SV-out}
    \widehat{\mathbb Var }_{\mathrm {out}}
    = \frac{1}{N_{\mathrm {out}}-1}\sum_{q=1}^{N_{\mathrm {out}}}\left(\hat y(\vect x_{q})-\widehat{\frac{\partial\mathbb{P}}{\partial\theta}}\right)^2
\end{equation}
and $\widehat{\mathbb Var }_{\mathrm {in}}$ is the averaged inner-level sampling variance, i.e.
\begin{equation}\label{SV-in}
    \widehat{\mathbb Var }_{\mathrm {in}}
   =\frac{1}{N_{\mathrm {out}}}\sum_{q=1}^{N_{\mathrm {out}}}\left[\frac{1}{N_{\mathrm {in}}(N_{\mathrm {in}}-1)}\sum_{j=1}^{N_{\mathrm {in}}}\left(\frac{z_{k_ji_j}(\vect x_{q})}{h(k_j,i_j)}-\hat y(\vect x_{q})\right)^2\right]
\end{equation}
Once the sample estimates of the mean and the variance are available, the sample estimate of the COV of the estimator can then be readily obtained as
\begin{equation}\label{COV}
    \widehat{\mathbb Cov }\!\left[\widehat{\frac{\partial\mathbb{P}}{\partial\theta}}\right]
   =\sqrt{\widehat{\mathbb Var }\!\left[\widehat{\frac{\partial\mathbb{P}}{\partial\theta}}\right]}  \bigg/ \left|\widehat{\frac{\partial\mathbb{P}}{\partial\theta}}\right|
\end{equation}
which is adopted as the stopping criterion for the proposed nested sampling algorithm.

\subsection{Solution procedure}
For clarity, a detailed solution procedure of the proposed surface decomposition method for the first-passage dynamic reliability sensitivity analysis of linear systems subjected to non-Gaussian wind excitations is given as follows:

i) Perform $N_v$ impulse response time-history analyses of the linear system, and obtain the coefficients $a^{s_{k,j}}_{i,1},a^{s_{k,j}}_{i,2},\dots,a^{s_{k,j}}_{i,i}\;(i=1,2,\dots,n;j=1,2,\dots,N_v)$ for all the responses of interest $s_k(\vect X,t)\;(k=1,2,\dots,m)$. Construct $\vect a_i^{s_k}$ and $\vect A_i^{s_k} \;(i=1,2,\dots,n;k=1,2,\dots,m)$ by Eqs.\eqref{vect a} and \eqref{matrix A}, respectively.

ii) For each design parameter $\theta$, perform $N_v$ impulse response sensitivity time-history analyses and obtain the coefficients $b^{s_{k,j}}_{i,1},b^{s_{k,j}}_{i,2},\dots,b^{s_{k,j}}_{i,i}\;(i=1,2,\dots,n;j=1,2,\dots,N_v)$ for the response sensitivities $\partial s_k(\vect X,t)/\partial \theta \;(k=1,2,\dots,m)$. Construct $\vect b_i^{s_k}$ and $\vect B_i^{s_k}\;(i=1,2,\dots,n;k=1,2,\dots,m)$ by Eqs.\eqref{vect b} and \eqref{matrix B}, respectively.

iii) Solve Eq.\eqref{optimization} for the design point $\vect{x}_{ki}^{\star}$ for all component limit-state functions $g_{ki}(\vect x)\;(i=1,2,\dots,n;k=1,2,\dots,m)$, and obtain the corresponding unit vectors $\vect u_{ki}$, reliability indexes $\beta_{ki}$, and the approximate failure probabilities $\tilde{\mathbb{P}}_{ki}$ by Eqs.\eqref{unit vector}, \eqref{beta}, and \eqref{component probability approximate}, respectively.

iv) Construct the importance sampling PMF $h(k,i)$ by Eq.\eqref{constructed ID approximate}.

v) Perform Algorithm \ref{algorithm} and obtain the estimate for the system failure probability sensitivity $\widehat{\partial\mathbb{P}}/\partial \theta$ for each design parameter $\theta$.

\begin{algorithm}[ht]
\caption{Nested Sampling Algorithm for Surface Decomposition Method} 
\label{algorithm}
\KwIn{$\vect a_i^{s_k}$ and $\vect A_i^{s_k}$, $\vect b_i^{s_k}$ and $\vect B_i^{s_k}$ for each $\theta$, $\vect u_{ki}$, $\beta_{ki}$, $\tilde{\mathbb{P}}_{ki}$, $h(k,i)$, $N_{\mathrm {in}}=100$, $N^{\min}_{\mathrm {out}}=10$, $N^{\max}_{\mathrm {out}}=10^3$, $tol=0.1$}
\BlankLine
\BlankLine
\KwOut{$\widehat{\partial\mathbb{P}}/\partial \theta$ for each $\theta$, $N_{\mathrm {out}}$}
\BlankLine
\For{$q \leftarrow 1$ \KwTo $N^{\max}_{\mathrm{out}}$}{
    Generate sample $\vect x_q \sim f_{\vect X} (\vect x)$\;
    Compute limit-state functions $g_{ki}(\vect x_q)\;(i=1,2,\dots,n;k=1,2,\dots,m)$\;
    Solve Eq.\eqref{quadratic equation} for $\Delta p_{ki}\;(i=1,2,\dots,n;k=1,2,\dots,m)$\;
    Generate index pairs $(k_j,i_j) \sim h(k,i)\;(j=1,2,\dots,N_{\mathrm{in}})$\;
    Compute 
    $\vect x_{k_ji_j} \leftarrow 
    \vect x_q+\Delta p_{k_ji_j}\vect u_{k_ji_j}\;(j=1,2,\dots,N_{\mathrm{in}})$\;
    Compute $p_{k_ji_j} \leftarrow \vect u_{k_ji_j}^{\mathrm T}\vect x_q+\Delta p_{k_ji_j}\;(j=1,2,\dots,N_{\mathrm{in}})$\;
    Compute indicator functions $\mathbb{I}_{k_j i_j}(\vect x_{k_ji_j})\;(j=1,2,\dots,N_{\mathrm{in}})$ by Eq.\eqref{indicator1}\;
    Compute $\nabla _{\vect x}g_{k_ji_j}(\vect x_{k_ji_j})\;(j=1,2,\dots,N_{\mathrm{in}})$ by Eq.\eqref{gradient}\;
    For each $\theta$, \\
    \quad \quad \quad \quad Compute $\partial g_{k_ji_j}(\vect x_{k_ji_j})/\partial \theta\;(j=1,2,\dots,N_{\mathrm{in}})$ by Eq.\eqref{sensitivity of component lsf1}\;
    \quad \quad \quad \quad Compute $z_{k_ji_j}(\vect x_{q})\;(j=1,2,\dots,N_{\mathrm{in}})$ by Eq.\eqref{z}\;
    \quad \quad \quad \quad Compute $\hat y(\vect x_{q})$ by Eq.\eqref{estimator2}\;
    \quad \quad \quad \quad Compute $\widehat{\partial\mathbb{P}}/\partial \theta$ by Eq.\eqref{estimator1}\;
    \quad \quad \quad \quad Compute $\widehat{\mathbb Var }[\widehat{\partial\mathbb{P}}/\partial \theta]$ by Eqs.\eqref{estimator variance approx}, \eqref{SV-out} and \eqref{SV-in}\;
    \quad \quad \quad \quad Compute $\widehat{\mathbb Cov }[\widehat{\partial\mathbb{P}}/\partial \theta]$ by Eq.\eqref{COV}\;
    Break the loop when $ \widehat{\mathbb Cov }[\widehat{\partial\mathbb{P}}/\partial \theta]<tol$  for each $\theta$ and $q>N_{\mathrm{out}}^{\min}$;
}
\BlankLine
\Return{$\widehat{\partial\mathbb{P}}/\partial \theta\; \mathrm{for\; each}\; \theta$, $N_{\mathrm{out}}=q$}
\end{algorithm}

It is noted that, when a large number of design parameters are considered, the adjoint variable method developed in \cite{hu2016explicit} is utilized to replace the numerous impulse response sensitivity analyses involved in Step ii) with merely a few impulse response analyses based on the adjoint equations. As a result, the computational cost  increases only marginally as the number of design parameters grows. Furthermore, it can be observed from Algorithm \ref{algorithm} that the dominant computational cost, namely, the evaluation of the limit-state functions $g_{ki}(\vect x_q)\;(i=1,2,\dots,n;k=1,2,\dots,m)$, is independent of the number of design parameters. Therefore, similar to the surface decomposition method originally developed for Gaussian excitations \cite{xian2026surface}, the present non-Gaussian extension maintains the same key advantage that its major computational expense does not scale with the number of design parameters.

\section{Numerical examples}
In this section, the effectiveness of the proposed surface decomposition method (SDM) for first-passage dynamic reliability sensitivity analysis of linear systems subjected to non-Gaussian wind excitations is validated using two numerical examples, namely, an oscillator and a 20-story planar braced frame structure. The domain decomposition method (DDM) developed in \cite{Wang2016efficient} under non-Gaussian wind excitations is employed to conduct the first-passage dynamic reliability analysis. Then, the finite difference method in conjunction with the above DDM with a sample size of $10^5$, denoted as FDM-DDM, is adopted to compute the reference solutions to the failure probability sensitivities. 

\subsection{An oscillator}\label{exam 1}
Consider the following equation of motion for a linear oscillator as 
\begin{equation}\label{oscillator equation}
     \ddot{u}(\vect X ,t)+2\omega_{n}\zeta_{n}  \dot{u}(\vect X ,t)+\omega^2_n u(\vect X ,t)
    = \lambda_1 v(\vect X ,t)+\lambda_2 v^2(\vect X ,t) 
\end{equation}
where $\omega_n=4\pi\mathrm{rad/s}$ and $\zeta_n=0.05$ denote the circular frequency and damping ratio, respectively; $\ddot{ u}(\vect X ,t)$, $\dot{u}(\vect X ,t)$ and $u(\vect X ,t)$ denote the acceleration, velocity and displacement, respectively; $v(\vect X ,t)$ is the fluctuating wind velocity, modeled as a zero-mean stationary Gaussian white noise with power spectral density $S_0=5.5\times 10^{-4}\mathrm{m}^2/\mathrm{s}$; and $\lambda_1$ and $\lambda_2$ are the coefficients associated with the linear and quadratic Gaussian components, respectively. The linear coefficient is fixed at $\lambda_1=1$, while the quadratic coefficient $\lambda_2$ is assigned with distinct values to represent varying degrees of non-Gaussianity in the wind excitation. For the spectral representation of $v(\vect X ,t)$, the upper and lower cutoff circular frequencies are set to $\omega_{\max}=25\pi\mathrm{rad/s}$ and $\omega_{\min}=0$, respectively, and the number of circular frequency intervals is chosen as $N_{\omega}=500$, resulting in a 1000-dimensional standard Gaussian random vector $\vect X$.

The failure event $F$ is defined as that the maximum displacement of the oscillator exceeds a prescribed threshold within a given time interval, which is expressed as
\begin{equation}\label{example1}
     F=\left\{c- \max_{t\in[0,T]} u(\vect X ,t) \leq 0\right\}
\end{equation}
where $c$ denotes the threshold, which considers different values to achieve varying levels of failure probability; and $T=n\Delta t=20\mathrm{s}$ is the time duration, with time step $\Delta t=0.02\mathrm{s}$ and $n=1000$ steps. Consequently, a total of 1000 component failure events, each corresponding to a discrete time instant, are taken into account.

Six different combinations of the quadratic coefficient $\lambda_2$ and the threshold $c$ are considered, which leads to six analysis cases: i) $\lambda_2=0$, $c=0.018\mathrm{m}$; ii) $\lambda_2=0.2$, $c=0.018\mathrm{m}$; iii) $\lambda_2=0.5$, $c=0.018\mathrm{m}$; iv) $\lambda_2=0.8$, $c=0.018\mathrm{m}$; v) $\lambda_2=0.8$, $c=0.016\mathrm{m}$; vi) $\lambda_2=0.8$, $c=0.014\mathrm{m}$. Note that Case i) corresponds to the Gaussian case where the quadratic Gaussian component is completely omitted. The results of failure probabilities for these cases, obtained from DDM with a target COV of 0.1, are presented in Table~\ref{tab:failure_probability Example 1}. By comparing the first four cases, it can be seen that the estimated failure probability increases with the degree of non-Gaussianity, suggesting that neglecting the quadratic Gaussian component may result in a significant underestimation of the failure probability. Furthermore, from the last three cases, it is observed that the efficiency of DDM mildly decreases as the failure probability becomes larger.

\begin{table}[htbp]
\centering
\footnotesize
\caption{Results of failure probabilities for Example~1.}
\label{tab:failure_probability Example 1}
\renewcommand{\arraystretch}{1.2}
\begin{tabular}{cccccc}
\hline
Case & $\lambda_2$ & $c$ (m) & $\mathbb{P}$ & Function Evaluations & COV \\
\hline
i) & 0 & 0.018 & $1.68\times10^{-7}$ & 13 & 0.0925 \\
ii) & 0.2 & 0.018 & $5.28\times10^{-7}$ & 25 & 0.0976 \\
iii) & 0.5 & 0.018 & $6.17\times10^{-6}$ & 29 & 0.0996 \\
iv) & 0.8 & 0.018 & $5.77\times10^{-5}$ & 35 & 0.0993 \\
v) & 0.8 & 0.016 & $6.68\times10^{-4}$ & 43 & 0.0997 \\
vi) & 0.8 & 0.014 & $7.17\times10^{-3}$ & 52 & 0.0995 \\
\hline
\end{tabular}
\end{table}

\begin{figure}[htbp]
  \centering
  \includegraphics[width=0.7\textwidth] {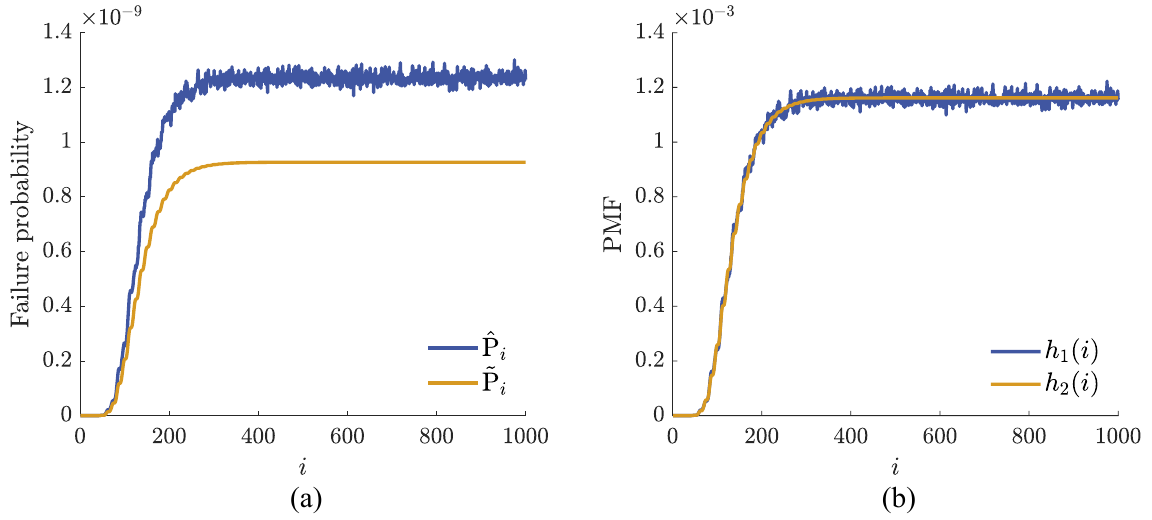}
  \caption{\color{Black}\textbf{Comparison between different importance sampling PMFs.} (a) Component failure probabilities estimated by line sampling and their first-order approximations; (b) Comparison between two PMFs.}  
  \label{fig:3}
\end{figure}

Before performing the first-passage reliability sensitivity analysis using SDM, the difference between the importance sampling PMFs constructed based on Eqs.\eqref{constructed ID} and \eqref{constructed ID approximate} is investigated. Case ii) is taken as an illustrative example. The component failure probabilities estimated using the line sampling with a sample size of 1000, which are denoted as $\hat{\mathbb{P}}_i\;(i=1,2,\dots,1000)$, along with the first-order approximations $\tilde{\mathbb{P}}_i\;(i=1,2,\dots,1000)$, are presented in Figure \ref{fig:3}(a). On this basis, two importance sampling PMFs, namely, $h_1(i)= \hat{\mathbb{P}}_i /\sum_{j=1}^{1000}\hat{\mathbb{P}}_j$ and $h_2(i)= \tilde{\mathbb{P}}_i /\sum_{j=1}^{1000}\tilde{\mathbb{P}}_j$, are constructed and depicted in Figure \ref{fig:3}(b). It is observed that, although the unbiased estimates of the component failure probabilities and their first-order approximations differ considerably, the resulting PMFs are in very good agreement. In fact, the PMF derived from the first-order approximations is arguably preferable, since it is purely analytical and therefore free from the sampling variability inherent in the line-sampling estimates.

\begin{figure}[htbp]
  \centering
  \includegraphics[width=1\textwidth] {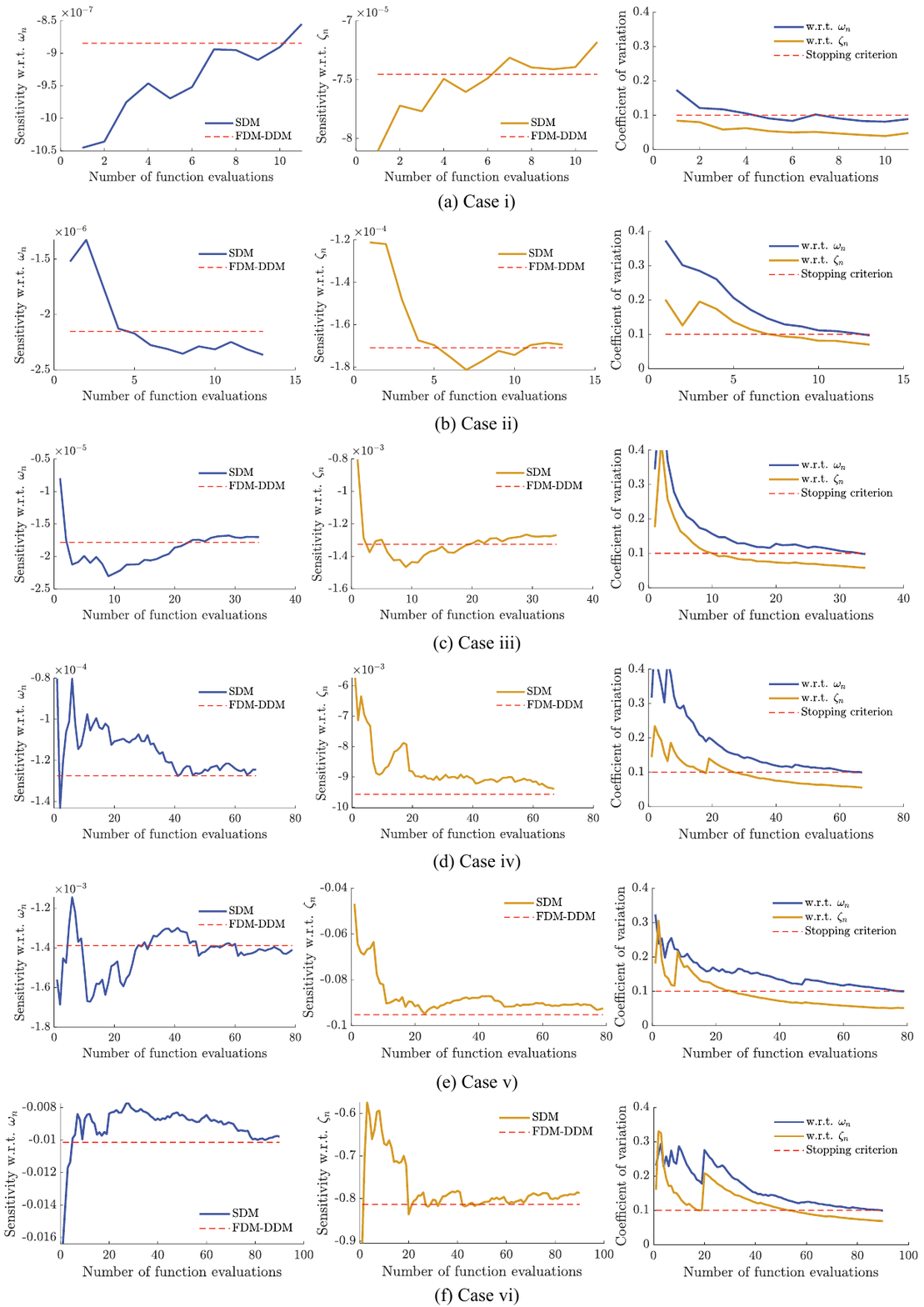}
  \caption{\color{Black}\textbf{Convergence histories of sensitivities of failure probability with respect to $\omega_n$ and $\zeta_n$ for Example 1.} The numbers of function evaluations required in SDM for Case i) to Case vi) are 11, 13, 34, 67, 79, and 90, respectively, to achieve a target COV of 0.1.}  
  \label{fig:4}
\end{figure}

The proposed SDM with a target COV of 0.1 is adopted to compute the sensitivities of the failure probability with respect to the circular frequency $\omega_n$ and damping ratio $\zeta_n$ of the oscillator. The sensitivity results for all the six cases are presented in Table~\ref{tab:failure_sensitivity for example 1}, and the corresponding convergence histories are depicted in Figure \ref{fig:4}. It is seen that the failure probability sensitivities obtained by SDM agree well with the reference solutions derived from FDM-DDM. 
Across all six cases, the number of evaluations of the system limit-state function required by SDM is fewer than 100, demonstrating the remarkable  computational efficiency of the proposed method. Moreover, a comparison of the first four cases indicates that increasing the degree of non-Gaussianity not only influences the failure probability but also significantly affects its sensitivities. Further, the comparison among the last three cases demonstrates that the efficiency of SDM improves as the failure probability decreases, which is consistent with the trend previously observed for Gaussian excitations in \cite{xian2026surface}.

\begin{table}[htbp]
\centering
\footnotesize
\caption{Results of sensitivities of failure probability with respect to $\omega_n$ and $\zeta_n$ for Example~1.}
\label{tab:failure_sensitivity for example 1}
\renewcommand{\arraystretch}{1.2}
\setlength{\tabcolsep}{4pt}
\begin{tabular}{c c c c c c c c c}
\hline
Case & $\lambda_2$  & $c$ (m) & Method & $\partial \mathbb{P}/\partial\omega_n$ & COV  & $\partial \mathbb{P}/\partial\zeta_n$ & COV & Function Evaluations \\
\hline
\multirow{2}{*}{i)} & \multirow{2}{*}{0} & \multirow{2}{*}{0.018} & SDM & $-8.55\times10^{-7}$ & 0.0888 & $-7.18\times10^{-5}$ & 0.0480 & 11 \\
 & & & FDM-DDM & $-8.84\times10^{-7}$ & —— & $-7.46\times10^{-5}$ & —— & $3\times10^{5}$ \\
\hline
\multirow{2}{*}{ii)} & \multirow{2}{*}{0.2} & \multirow{2}{*}{0.018} & SDM & $-2.36\times10^{-6}$ & 0.0971 & $-1.69\times10^{-4}$ & 0.0699 & 13 \\
 & & & FDM-DDM & $-2.15\times10^{-6}$ & —— & $-1.71\times10^{-4}$ & —— & $3\times10^{5}$ \\
\hline
\multirow{2}{*}{iii)} & \multirow{2}{*}{0.5} & \multirow{2}{*}{0.018} & SDM & $-1.70\times10^{-5}$ & 0.0974 & $-1.27\times10^{-3}$ & 0.0577 & 34 \\
 & & & FDM-DDM & $-1.78\times10^{-5}$ & —— & $-1.33\times10^{-3}$ & —— & $3\times10^{5}$ \\
\hline
\multirow{2}{*}{iv)} & \multirow{2}{*}{0.8} & \multirow{2}{*}{0.018} & SDM & $-1.25\times10^{-4}$ & 0.0995 & $-9.40\times10^{-3}$ & 0.0555 & 67 \\
 & & & FDM-DDM & $-1.28\times10^{-4}$ & —— & $-9.57\times10^{-3}$ & —— & $3\times10^{5}$ \\
\hline
\multirow{2}{*}{v)} & \multirow{2}{*}{0.8} & \multirow{2}{*}{0.016} & SDM & $-1.41\times10^{-3}$ & 0.1000 & $-9.25\times10^{-2}$ & 0.0511 & 79 \\
 & & & FDM-DDM & $-1.39\times10^{-3}$ & —— & $-9.53\times10^{-2}$ & —— & $3\times10^{5}$ \\
\hline
\multirow{2}{*}{vi)} & \multirow{2}{*}{0.8} & \multirow{2}{*}{0.014} & SDM & $-9.79\times10^{-3}$ & 0.0989 & $-7.87\times10^{-1}$ & 0.0681 & 90 \\
 & & & FDM-DDM & $-1.01\times10^{-2}$ & —— & $-8.14\times10^{-1}$ & —— & $3\times10^{5}$ \\
\hline
\end{tabular}
\end{table}

\subsection{A 20-story planar braced frame structure}
\begin{figure}
  \centering  \includegraphics[width=0.8\textwidth] {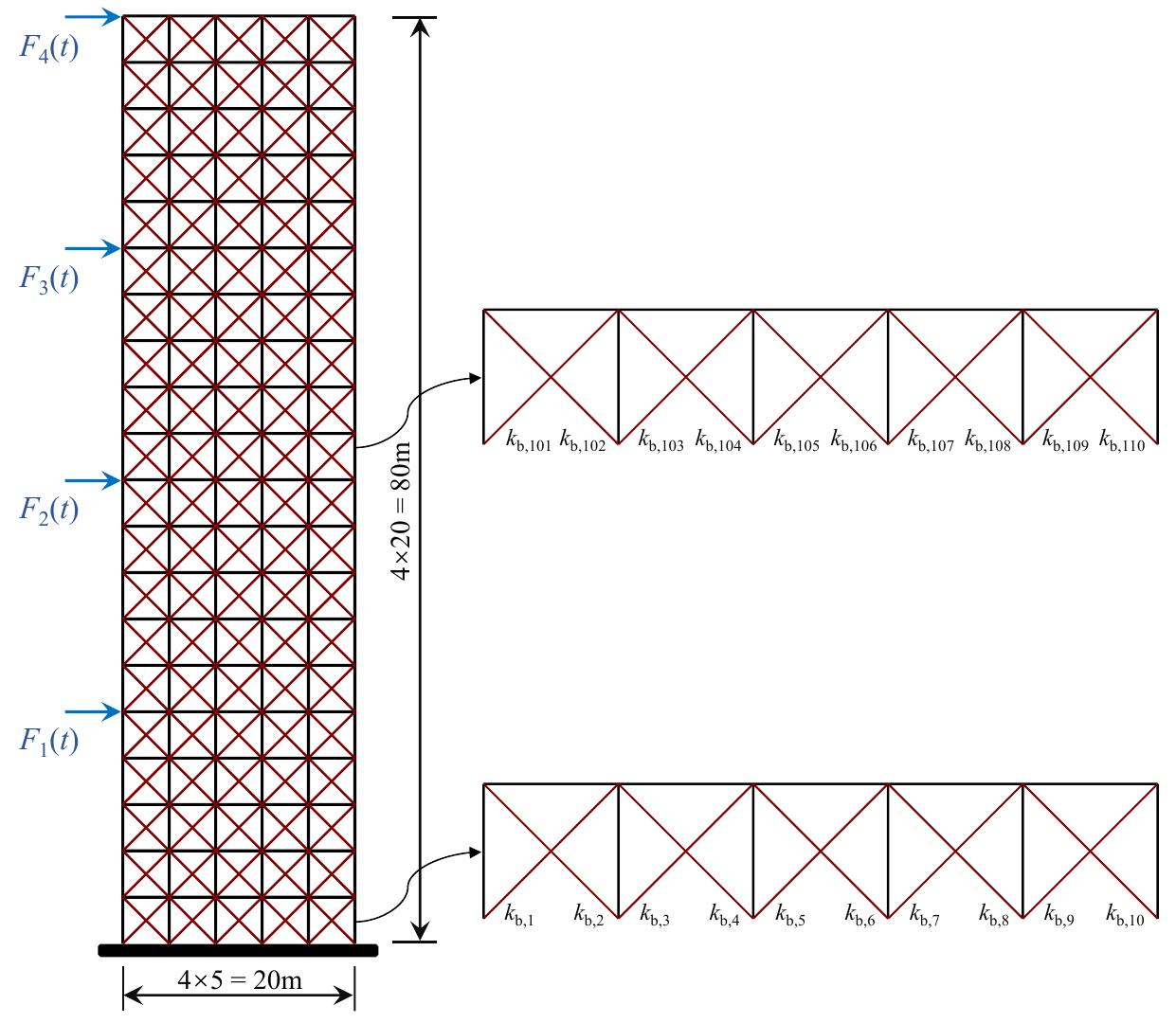}
  \caption{\color{Black}\textbf{A 5-bay 20-story planar braced frame structure.} }  
  \label{fig:5}
\end{figure}

Consider a 5-bay 20-story planar reinforced concrete
frame structure equipped with steel braces, as shown in Figure \ref{fig:5}. Each bay has a span of $4\mathrm m$ and each story has a height of $4\mathrm m$, resulting in a total width of $20\mathrm m$ and a total height of $80\mathrm m$. The cross section of each frame member is $0.5\mathrm m \times 0.5\mathrm m$. The Young’s modulus and mass density of all frame members are $E_0=20\mathrm {GPa}$ and $\rho_0 = 2500\mathrm {kg/m^3}$, respectively, and the mass of the frame member is lumped at the beam-column nodes. Each story contains 10 steel braces, which leads to a total of 200 braces installed in the structure. The numbering of these steel braces is illustrated in Figure \ref{fig:5}, and the stiffness of each brace is assumed as $k_{\mathrm b,i}=k_{\mathrm b}=4500\mathrm{kN/m}\;(i=1,2,\dots,200)$.
The frame members are modeled using beam elements, while the steel braces are simulated by truss elements, yielding a total of 360 degrees of freedom. The Rayleigh damping model is adopted to define the damping matrix, with a damping ratio of 0.05 assigned to the 1st and 360th modes of the structure.

The braced frame structure is subjected to along-wind excitations. The wind velocity field is discretized into $V_j(t)\;(j=1,2,3,4)$ at heights $h_1=20\mathrm m$, $h_2=40\mathrm m$, $h_3=60\mathrm m$, and $h_4=80\mathrm m$, respectively, producing four concentrated wind loads $F_j(t)\;(j=1,2,3,4)$ acting on the structure, as illustrated in Figure \ref{fig:5}. The effective area for the wind pressure from $V_j(t)$ is $S_j=100\mathrm m^2\;(j=1,2,3,4)$. The mean component of $V_j(t)$ follows the power law below \cite{simiu2019wind}:
\begin{equation}\label{mean wind speed}
\bar V_j=41\left( \frac{h_j}{180}\right)^{0.25}\quad (j=1,2,3,4)
\end{equation}
With $S_j$ and $\bar V_j$, the static wind load $\bar{F}_j$ along with the linear and quadratic coefficients $\lambda_{j,1}$ and $\lambda_{j,2}\;(j=1,2,3,4)$ in Eq.\eqref{wind load2} can be readily determined.

The fluctuating component of $V_j(t)$ is assumed to be a zero-mean uniformly modulated non-stationary Gaussian random process, namely, $v_j(t)=g_j(t)v^0_j(t)$, in which $v^0_j(t)$ is the underlying stationary part, and $g_j(t)$ is the modulation function expressed as
\begin{equation}\label{modulation function}
g_j(t) = g(t)= \beta_1 t^{\beta_2} \exp(-\beta_3 t)\quad(j=1,2,3,4)
\end{equation}
where $\beta_1=0.0012$, $\beta_2=2$, $\beta_3=0.025$.

The auto-power spectrum of $v^0_j(t)$ is given by \cite{davenport1962buffeting}
\begin{equation}\label{auto PSD}
S_{jj}^0(\omega)= \frac{\bar{V}_j^2 K^2}{\left(\ln (h_j/h_0)\right)^2} \frac{8 \pi \chi^2}{\omega (1 + \chi^2)^{4/3}} \quad(j=1,2,3,4)
\end{equation}
where $K=0.4$ is the Von Karman’s constant; $h_0=0.05\mathrm m$ is the roughness length; and 
\begin{equation}\label{chi}
\chi= \frac{600\omega}{ \pi \bar V(10)}
\end{equation}
where $\bar V(10)=19.9\mathrm{m/s^2}$ is the mean wind speed at the height of 10m.

The cross-power spectrum of $v^0_j(t)$ and $v^0_k(t)$ is given by
\begin{equation}\label{cross PSD}
S_{jk}^0(\omega)= \sqrt{S_{jj}^0(\omega)S_{kk}^0(\omega)} \gamma_{jk}(\omega) \quad(j,k=1,2,3,4)
\end{equation}
where $\gamma_{jk}(\omega)$ is the coherence function between $v^0_j(t)$ and $v^0_k(t)$, namely \cite{davenport1962buffeting}
\begin{equation}\label{coherence}
\gamma_{jk}(\omega)= \exp\! \left[ - \frac{\omega C_h |h_j - h_k|}{ \pi( \bar{V}_j + \bar{V}_k )} \right] \quad(j,k=1,2,3,4)
\end{equation}
where $C_h=10$ is the decay coefficient for the vertical coherence.

For the spectral representation of $v_j(\vect X,t)\;(j=1,2,3,4)$, the upper and lower cutoff circular frequencies are $\omega_{\max}=\pi\mathrm{rad/s}$ and $\omega_{\min}=0$, respectively, and the width of each circular frequency interval is set to be $\Delta \omega=2\pi/T$, with $T=400\mathrm s$ being the time duration of the wind excitations. For $N_v=4$ wind velocity processes and $N_{\omega}=200$ circular frequency intervals, the dimension of the standard Gaussian random vector $\vect X$ is $d=2N_vN_{\omega}=1600$.

The failure event $F$ is defined as that the maximum value of the top-story drift $u_{\mathrm{top}}(\vect X ,t)$ of the braced frame structure exceeds a given threshold $c=0.2\mathrm{m}$ over the time duration $T=400\mathrm{s}$, which is expressed as
\begin{equation}\label{example2}
     F=\left\{c-  \max_{t\in[0,T]}  u_{\mathrm{top}}(\vect X ,t) \leq 0\right\}
\end{equation}
The time step is set as $\Delta t=0.05\mathrm s$, and thus the number of time steps is $n=8000$, resulting in a total of 8000 component failure events.

The failure probability of the braced frame structure subjected to non-Gaussian wind excitations is determined to be $4.47\times10^{-3}$ using DDM, with 49 evaluations of the system limit-state function required to achieve a target COV of 0.1. In contrast, when the quadratic Gaussian components of the wind excitations are neglected, which renders the excitations purely Gaussian, the corresponding failure probability drops sharply to $2.95\times 10^{-7}$. This stark contrast indicates that omitting the quadratic terms leads to a severe underestimation of the failure probability by approximately four orders of magnitude, and consequently, a significant overestimation of the structural safety.

\begin{figure}[htbp]
  \centering
  \includegraphics[width=1\textwidth] {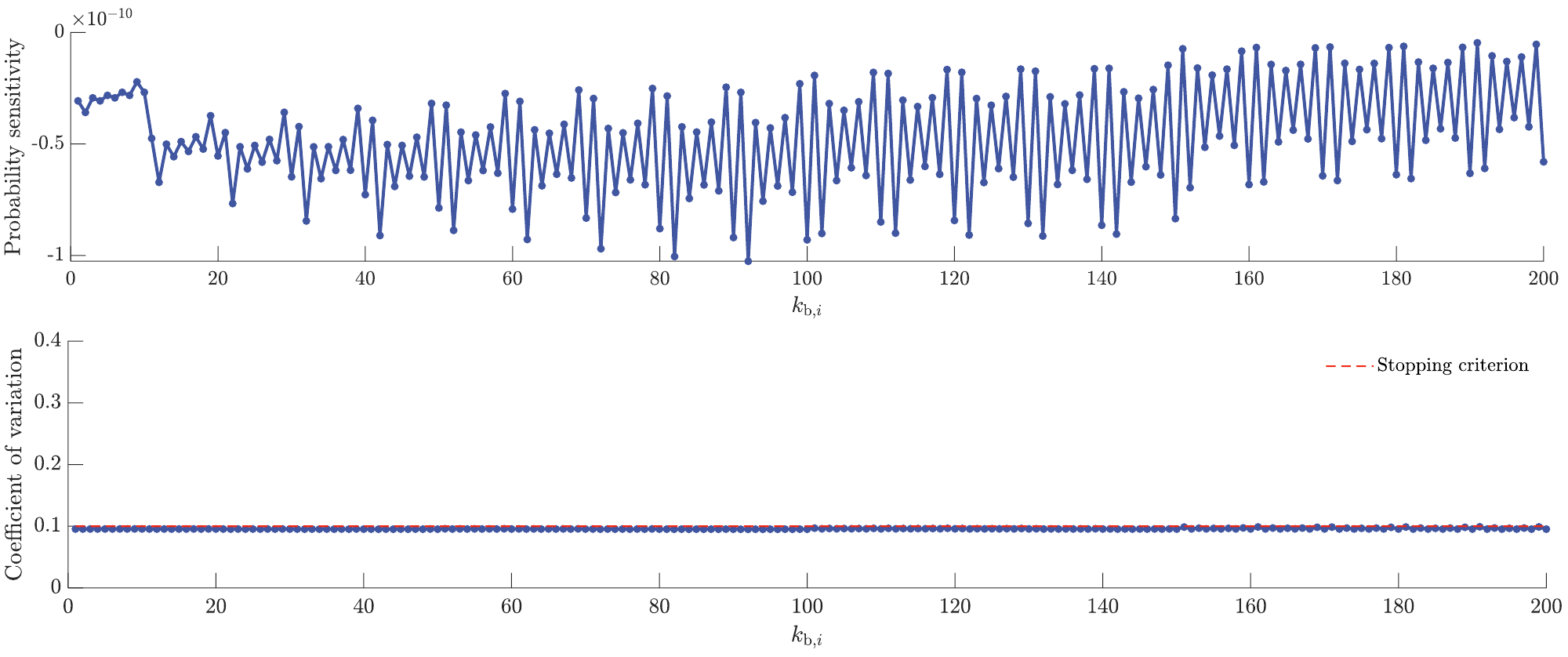}
  \caption{\color{Black}\textbf{Results of failure probability sensitivities with respect to $k_{\mathrm b,i}\;(i=1,2,\dots,200)$ for Example 2.} The number of function evaluations required in SDM is 51 for a target COV of 0.1.}  
  \label{fig:6}
\end{figure}

\begin{figure}[htbp]
  \centering
  \includegraphics[width=0.8\textwidth] {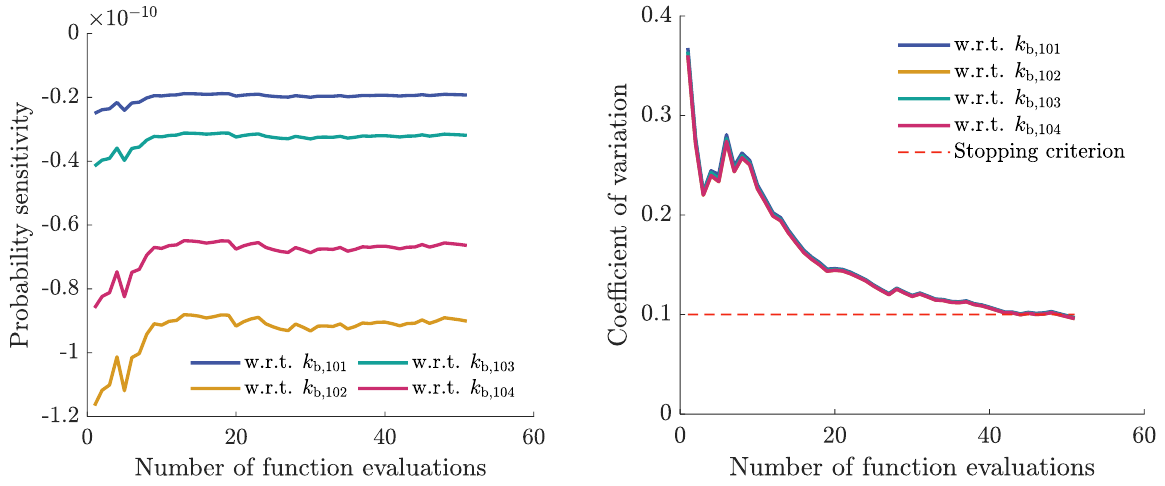}
  \caption{\color{Black}\textbf{Convergence histories of failure probability sensitivities with respect to $k_{\mathrm b,101}$, $k_{\mathrm b,102}$, $k_{\mathrm b,103}$, and $k_{\mathrm b,104}$ for Example 2.} }  
  \label{fig:7}
\end{figure}

The SDM is then employed for the sensitivity analysis of the first-passage dynamic reliability considering non-Gaussian wind excitations, where the stiffnesses of all the steel braces, namely, $k_{\mathrm b,i}\;(i=1,2,\dots,200)$, are taken as the design parameters.
The results of the failure probability sensitivities with respect to all the design parameters are depicted in Figure \ref{fig:6}, where the COVs of these sensitivity estimates are all below 0.1. Moreover, the convergence histories of the failure probability sensitivities with respect to $k_{\mathrm b,101}$, $k_{\mathrm b,102}$, $k_{\mathrm b,103}$, and $k_{\mathrm b,104}$ are presented in Figure \ref{fig:7}. It is observed that increasing the stiffness of any steel brace helps reduce the failure probability, and the failure probability is most sensitive to $k_{\mathrm b,92}$ and least sensitive to $k_{\mathrm b,191}$. The sensitivity results can be further utilized to guide the design optimization of the parameters and/or layouts of the steel braces in conjunction with gradient-based optimizers. Importantly, the computation of all 200 sensitivity estimates requires only 51 evaluations of the system limit-state function, demonstrating remarkable computational efficiency of SDM and its great potential for reliability-based design optimization.

\section{Conclusions}
A surface decomposition method has been developed for the sensitivity analysis of first-passage dynamic reliability of linear structures subjected to non-Gaussian wind excitations. The method features a decomposition of the original highly non-smooth system limit-state hypersurface into a non-overlapping collection of truncated smooth quadratic component limit-state hypersurfaces. Accordingly, the complex system surface integral, which represents the system failure probability sensitivity, is decomposed into a sum of component surface integrals that are substantially easier to evaluate. Mathematically, the core concept of the proposed method is to convert a challenging non-smooth integration problem into a series of more tractable smooth integration subproblems.
Following surface decomposition, an importance sampling is introduced to identify the dominant component surface integrals based on the relative magnitudes of the first-order approximations of the component failure probabilities. A two-stage nested sampling algorithm is further developed to efficiently estimate the total contribution of the above component surface integrals. The high efficiency stems from two aspects: i) the number of evaluations of the system limit-state function is only associated with the outer-level sample size and is independent of the inner-level sample size; ii) closed-form expressions are available for the component and system limit-state functions as well as their sensitivities for linear systems under non-Gaussian wind excitations.

The effectiveness of the surface decomposition method has been validated using two numerical examples, including a linear oscillator and a 20-story planar braced frame structure. It has been found that neglecting the quadratic components of the non-Gaussian wind excitations results in a severe underestimation of the failure probability and also substantially affects the corresponding failure probability sensitivities. Furthermore, the proposed method becomes increasingly efficient as the failure probability decreases. Its major computational cost is essentially independent of the number of design parameters, since the system limit-state function evaluations can be reused for the sensitivity analyses with respect to different design parameters. For all the cases investigated in this study, only fewer than 100 system limit-state function evaluations are required to achieve a target COV of 0.1, even when the number of design variables reaches 200 in the second example. These results demonstrate the excellent computational efficiency of the proposed method as well as its strong potential for reliability-based design optimization.

\section{Acknowledgments}
This work was supported by the start-up fund and the seed fund for Basic Research for New Staff from the University of Hong Kong. The first and corresponding authors gratefully acknowledge the support of the University of Hong Kong.

\bibliography{Ref}

\appendix
\section{Derivation of Eqs.\eqref{component surface integral LS} and \eqref{z} using line sampling }\label{appendix a}
The component surface integral $\eta_{ki}$ in Eq.\eqref{component surface integral} is evaluated using line sampling. Specifically, any point on the component limit-state hypersurface $S_{ki}$ can be expressed as
\begin{equation}\label{A1}
    \vect x_{ki}
    = \vect x^\perp_{ki} + p_{ki}\vect u_{ki}\quad(i=1,2,\dots,n;k=1,2,\dots,m)
\end{equation}
where $\vect u_{ki}$ is the sampling direction vector shown in Eq.\eqref{unit vector}; $\vect x^\perp_{ki}=\vect x-(\vect u_{ki}^{\mathrm T}\vect x)\vect u_{ki}$ denotes the projection of $\vect x$ onto an auxiliary hyperplane passing through the origin and perpendicular to $\vect u_{ki}$; and $p_{ki}$ is determined by solving 
$g_{ki}(\vect x^\perp_{ki}+p_{ki}\vect u_{ki})=0$.

The surface element $dS_{ki}$ on $S_{ki}$ is related to its projection $d\vect x^\perp_{ki}$ on the auxiliary hyperplane through
\begin{equation}\label{A2}
    d S_{ki}
    = \frac{\left\|\nabla _{\vect x}g_{ki}(\vect x_{ki})\right\|}{\left|\vect u_{ki}^{\mathrm T}\nabla _{\vect x}g_{ki}(\vect x_{ki})\right|} d\vect x^\perp_{ki} \quad(i=1,2,\dots,n;k=1,2,\dots,m)
\end{equation}
Substituting Eq.\eqref{A2} into Eq.\eqref{component surface integral} yields
\begin{equation}\label{A3}
    \eta_{ki}
    = -\int_{\vect x^\perp_{ki}} \frac{1}{\left|\vect u_{ki}^{\mathrm T}\nabla _{\vect x}g_{ki}(\vect x_{ki})\right|} \frac{\partial g_{ki}(\vect x_{ki})}{\partial \theta}\mathbb{I}_{ki}(\vect x_{ki})f_{\vect X}(\vect x_{ki})\,  d\vect x^\perp_{ki} \quad(i=1,2,\dots,n;k=1,2,\dots,m)
\end{equation}

Since the $(d-1)$-dimensional random vector $\vect X^\perp_{ki}$ and the one-dimensional component along $\vect u_{ki}$ constitute an orthogonal decomposition of the standard Gaussian random vector $\vect X$, these two components are mutually independent, and then the joint PDF $f_{\vect X}(\vect x_{ki})$ in Eq.\eqref{A3} can be factorized as
\begin{equation}\label{A4}
    f_{\vect X}(\vect x_{ki})
    = f_{\vect X}( \vect x^\perp_{ki} + p_{ki}\vect u_{ki})= f_{\vect X^\perp_{ki}}( \vect x^\perp_{ki})\varphi(p_{ki})\quad(i=1,2,\dots,n;k=1,2,\dots,m)
\end{equation}
where $\varphi (\cdot)$ denotes the PDF of the standard Gaussian distribution. The marginal PDF $f_{\vect X^\perp_{ki}}( \vect x^\perp_{ki})$ can be expressed as
\begin{equation}\label{A5}
    f_{\vect X^\perp_{ki}}( \vect x^\perp_{ki})
    =\int_{-\infty}^{\infty} f_{\vect X}(\vect x^\perp_{ki} + t\vect u_{ki})\,dt\quad(i=1,2,\dots,n;k=1,2,\dots,m)
\end{equation}

Then, substituting Eqs.\eqref{A4} and \eqref{A5} into Eq.\eqref{A3}, one has
\begin{equation}\label{A6}
    \eta_{ki}
    = -\displaystyle\int_{\mathbb{R}^d} \dfrac{1}{\left|\vect u_{ki}^{\mathrm T}\nabla _{\vect x}g_{ki}(\vect x_{ki})\right|} \dfrac{\partial g_{ki}(\vect x_{ki})}{\partial \theta}\mathbb{I}_{ki}(\vect x_{ki})\varphi (p_{ki})f_{\vect X}(\vect x)\, d \vect x \quad(i=1,2,\dots,n;k=1,2,\dots,m)
\end{equation}

\section{Mean and variance of the nested Monte Carlo estimator }\label{appendix b}
The nested Monte Carlo estimator defined in Eq.\eqref{estimator1} is a random variable denoted as
\begin{equation}\label{B1}
    Q=\widehat{\frac{\partial\mathbb P}{\partial\theta}}
    = \frac{1}{N_{\mathrm {out}}}\sum_{q=1}^{N_{\mathrm {out}}}\hat y(\vect X_{q})
\end{equation}
where
\begin{equation}\label{B2}
    \hat y(\vect X_{q})
   =\frac{1}{N_{\mathrm {in}}}\sum_{j=1}^{N_{\mathrm {in}}}\frac{z_{K_jI_j}(\vect X_{q})}{h(K_j,I_j)}
\end{equation}
Here, $\vect X_q\;(q=1,2,\dots,N_{\mathrm{out}})$ are mutually independent and identically distributed random vectors with PDF $f_{\vect X}(\vect x)$; and $(K_j,I_j)\;(j=1,2,\dots,N_{\mathrm{in}})$ are mutually independent and identically distributed random index pairs with PMF $h(k,i)$.

The mean and variance of $Q$ are given by
\begin{equation}\label{B3}
    \mathbb E[Q]= \frac{1}{N_{\mathrm {out}}}\sum_{q=1}^{N_{\mathrm {out}}}\mathbb E_{\vect X}[\hat y(\vect X_{q})] =\frac{1}{N_{\mathrm {out}}}N_{\mathrm {out}}\mathbb E_{\vect X}[\hat y(\vect X)]=\mathbb E_{\vect X}[\hat y(\vect X)]
\end{equation}
and
\begin{equation}\label{B4}
    \mathbb Var[Q]= \frac{1}{N_{\mathrm {out}}^2}\sum_{q=1}^{N_{\mathrm {out}}}\mathbb Var_{\vect X}[\hat y(\vect X_{q})]= \frac{1}{N_{\mathrm {out}}^2}N_{\mathrm {out}}\mathbb Var_{\vect X}[\hat y(\vect X)]=\frac{1}{N_{\mathrm {out}}} \mathbb Var_{\vect X}[\hat y(\vect X)]
\end{equation}
respectively. Then, applying the laws of total expectation and total variance to $\hat y(\vect X)$ yields
\begin{equation}\label{B5}
\mathbb E_{\vect X}[\hat y(\vect X)]
=\mathbb E_{\vect X}\!\left[\mathbb E_h\!\left(
\hat y(\vect X)|\vect X\right)\right]
\end{equation}
and
\begin{equation}\label{B6}
\mathbb Var_{\vect X}[\hat y(\vect X)]
=\mathbb Var_{\vect X}\!\left[
\mathbb E_h\!\left(\hat y(\vect X)|\vect X
\right)\right]+\mathbb E _{\vect X}\!\left[
\mathbb Var _h\!\left(\hat y(\vect X)|\vect X
\right)\right]
\end{equation}
respectively, where $\mathbb E_h\!\left(\hat y(\vect X)|\vect X
\right)$ and $\mathbb Var _h\!\left(\hat y(\vect X)|\vect X
\right)$ are the conditional expectation and variance of $\hat y(\vect X)$ given $\vect X$, respectively. 

From Eq.\eqref{B2}, the conditional expectation and variance can be derived as
\begin{equation}\label{B7}
    \mathbb E_h\!\left(
\hat y(\vect X)|\vect X\right)=\frac{1}{N_{\mathrm {in}}}\sum_{j=1}^{N_{\mathrm {in}}}\mathbb E_h \! \left (\frac{z_{K_jI_j}(\vect X)}{h(K_j,I_j)}\middle|\vect X\right)=\frac{1}{N_{\mathrm {in}}}N_{\mathrm {in}}\mathbb E_h \!\left (\frac{z_{KI}(\vect X)}{h(K,I)}\middle| \vect X\right)=y(\vect X)
\end{equation}
and
\begin{equation}\label{B8}
\begin{split}
    \mathbb Var_h\!\left(\hat y(\vect X)|\vect X
\right)&= \frac{1}{N_{\mathrm {in}}^2}\sum_{j=1}^{N_{\mathrm {in}}}\mathbb Var_h\!\left(\frac{z_{K_jI_j}(\vect X)}{h(K_j,I_j)}\middle|\vect X\right)= \frac{1}{N_{\mathrm {in}}^2} N_{\mathrm {in}} \mathbb Var_h \!\left(\frac{z_{KI}(\vect X)}{h(K,I)}\middle|\vect X\right)\\&= \frac{1}{N_{\mathrm {in}}} \mathbb Var_h \!\left(\frac{z_{KI}(\vect X)}{h(K,I)}\middle|\vect X\right)
\end{split}
\end{equation}
respectively. Substituting Eqs.\eqref{B7} and \eqref{B8} into Eqs.\eqref{B5} and \eqref{B6} yields the total expectation and variance as
\begin{equation}\label{B9}
\mathbb E_{\vect X}[\hat y(\vect X)]
=\mathbb E_{\vect X}\!\left[y(\vect X)\right]
\end{equation}
and
\begin{equation}\label{B10}
\mathbb Var_{\vect X}[\hat y(\vect X)]
=\mathbb Var_{\vect X}\!\left[y(\vect X)\right]+\mathbb E_{\vect X}\!\left[\frac{1}{N_{\mathrm {in}}} \mathbb Var_h \!\left(\frac{z_{KI}(\vect X)}{h(K,I)}\middle|\vect X\right)\right]
\end{equation}
respectively. 

Finally, substituting Eqs.\eqref{B9} and \eqref{B10} into Eqs.\eqref{B3} and \eqref{B4} leads to the mean and variance of $Q$ as
\begin{equation}\label{B11}
    \mathbb E[Q]=\mathbb E_{\vect X}\!\left[y(\vect X)\right]=\frac{\partial\mathbb P}{\partial\theta}
\end{equation}
and
\begin{equation}\label{B12}
    \mathbb Var[Q]=\frac{1}{N_{\mathrm {out}}} \left [ \mathbb Var_{\vect X}\!\left[y(\vect X)\right]+\mathbb E_{\vect X}\!\left[\frac{1}{N_{\mathrm {in}}} \mathbb Var_h \!\left(\frac{z_{KI}(\vect X)}{h(K,I)}\middle|\vect X\right)\right]\right]
\end{equation}
respectively. Eq.\eqref{B11} indicates that the nested Monte Carlo estimator is unbiased.

\end{document}